%% file: main.tex
\documentclass[sigplan,screen, table]{acmart}
\usepackage{amsmath,amsfonts}
\usepackage{algorithmic}
\usepackage{graphicx}
\usepackage{stfloats}
\usepackage{textcomp}
\usepackage{multirow}
\usepackage{multicol}
\usepackage{booktabs}
\usepackage{hhline}
\usepackage{bbding}
\usepackage[linesnumbered,ruled,vlined]{algorithm2e}
\usepackage{colortbl}
\usepackage{booktabs}
\usepackage{array}
\usepackage{enumitem}
\usepackage[misc]{ifsym}
\usepackage{xcolor}
\usepackage{subfigure}
\usepackage{tikz}
\newlength\szg     \newcommand\hquan[1]{%
\settoheight\szg{#1}%
\tikz[baseline]{
\pgfmathparse{1}
\let\hfs\pgfmathresult
\filldraw (0,\szg/2) circle (\szg/2+0.35ex);
\node[white] at (0,\szg/2) {\makebox[0em][c]{\scalebox{\hfs}[1]{\textbf{#1}}}};
}}
\definecolor{MintGreen}{HTML}{32CD32}
\definecolor{SalmonRed}{HTML}{F08080}
\definecolor{LightGray}{HTML}{DCDCDC}

\copyrightyear{2026}
\acmYear{2026}
\setcopyright{acmlicensed}
\acmConference[ASPLOS '26] {Proceedings of the 31st ACM International Conference on Architectural Support for Programming Languages and Operating Systems, Volume 1}{March 22--26, 2026}{Pittsburgh, PA, USA.}
\acmBooktitle{Proceedings of the 31st ACM International Conference on Architectural Support for Programming Languages and Operating Systems, Volume 1 (ASPLOS '26), March 22--26, 2026, Pittsburgh, PA, USA}
\acmISBN{979-8-4007-2165-6/2026/03}
\acmDOI{10.1145/3760250.3762229}
\begin{document}

\title{AGS: \underline{A}ccelerating 3D \underline{G}aussian Splatting \underline{S}LAM via CODEC-Assisted Frame Covisibility Detection}

\author{Houshu He}
\affiliation{School of Computer Science, Shanghai Jiao Tong University, Shanghai, China\country{}}
\email{h_houshu@sjtu.edu.cn}

\author{Naifeng Jing}
\affiliation{Department of Micro-Nano Electronics, Shanghai Jiao Tong University, Shanghai, China\country{}}
\email{sjtuj@sjtu.edu.cn}

\author{Li Jiang}
\affiliation{School of Computer Science, Shanghai Jiao Tong University, Shanghai, China\country{}}
\email{ljiang_cs@sjtu.edu.cn}

\author{Xiaoyao Liang}
\affiliation{School of Computer Science, Shanghai Jiao Tong University, Shanghai, China\country{}}
\email{liang-xy@cs.sjtu.edu.cn}

\author{Zhuoran Song}
\authornote{Zhuoran Song is the corresponding author.}
\affiliation{School of Computer Science, Shanghai Jiao Tong University, Shanghai, China\country{}}
\email{songzhuoran@sjtu.edu.cn}

\begin{abstract}
Simultaneous Localization and Mapping (SLAM) is a critical task that enables autonomous vehicles to construct maps and localize themselves in unknown environments. Recent breakthroughs combine SLAM with 3D Gaussian Splatting (3DGS) to achieve exceptional reconstruction fidelity. However, existing 3DGS-SLAM systems provide insufficient throughput due to the need for multiple training iterations per frame and the vast number of Gaussians.

In this paper, we propose AGS, an algorithm-hardware co-design framework to boost the efficiency of 3DGS-SLAM based on the intuition that SLAM systems process frames in a streaming manner, where adjacent frames exhibit high similarity that can be utilized for acceleration. On the software level: 1) We propose a coarse-then-fine-grained pose tracking method with respect to the robot's movement. 2) We avoid redundant computations of Gaussians by sharing their contribution information across frames. On the hardware level, we propose a frame covisibility detection engine to extract intermediate data from the video CODEC. We also implement a pose tracking engine and a mapping engine with workload schedulers to efficiently deploy the AGS algorithm. Our evaluation shows that AGS achieves up to $17.12\times$, $6.71\times$, and $5.41\times$ speedups against the mobile and high-end GPUs, and a state-of-the-art 3DGS accelerator, GSCore.
\end{abstract}


\maketitle

\input{Introduction}
\input{Background}
\input{Motivation}
\input{Algorithm}

\input{Architecture}
\input{Experiment}
\input{Related}
\input{Conclusion}

\bibliographystyle{ACM-Reference-Format}
\bibliography{main}

\end{document}

%% file: Introduction.tex
\section{Introduction}

Simultaneous Localization and Mapping (SLAM) is a pivotal task in computer vision, allowing autonomous systems such as robots to perceive and interact with the environment. The main objective of SLAM includes \textbf{tracking} and \textbf{mapping} that enable a robot to determine its own location and construct a map of the surroundings in an unknown environment. While both traditional~\cite{izadi2011kinectfusion, mur2015orb, mur2017orb, whelan2012kintinuous, whelan2015elasticfusion} and deep learning-based SLAM systems~\cite{li2020structure, tateno2017cnn, bloesch2018codeslam} struggle to capture fine details in 3D scenes, radiance field-based methods~\cite{li2021neural, mildenhall2021nerf, muller2022instant, kerbl20233d, barron2021mip}, represented by 3D Gaussian Splatting (3DGS)~\cite{kerbl20233d}, have demonstrated outstanding capabilities in reconstructing high-fidelity scenes, and are recently adopted in state-of-the-art SLAM systems~\cite{sucar2021imap, zhu2022nice, yan2024gs, matsuki2024gaussian, yang2022vox}. Table~\ref{tab_intro} compares the 3DGS-SLAM with Neural Radiance Field (NeRF)-based SLAM~\cite{mildenhall2021nerf, zhu2022nice, yang2022vox} and traditional SLAM~\cite{mur2017orb, whelan2012kintinuous}. While traditional SLAM algorithms deliver high tracking accuracy, they struggle to provide satisfying mapping results. In contrast, 3DGS-SLAM stands out for its sufficient tracking accuracy and superior mapping quality.

\begin{table}[!t]
\renewcommand\arraystretch{1.5}
\caption{Comparison of 3DGS-SLAM with existing SLAM algorithms. Green boxes highlight higher performance and red boxes indicate weaker performance.}
\vspace{-0pt}
\label{tab_intro}
\setlength{\arrayrulewidth}{0.83pt}
\definecolor{myred}{HTML}{EFC7C7}
\definecolor{mygreen}{HTML}{D5E4CE}
\resizebox{\columnwidth}{!}{
\begin{tabular}{c|c|c|c|c}
\hhline{-----}
\textbf{Category}           & \textbf{Algorithm}   & \textbf{\begin{tabular}[c]{@{}c@{}}Tracking Accuracy\\ ATE(cm)\end{tabular}} & \textbf{\begin{tabular}[c]{@{}c@{}}Mapping Accuracy\\ PSNR(dB)\end{tabular}} & \textbf{\begin{tabular}[c]{@{}c@{}}Speed Performance\\ Latency (s/frame)\end{tabular}} \\ \hhline{-----}
                            & SplatAM (2024)~\cite{keetha2024splatam}       & \cellcolor{myred!90}{\color[HTML]{000000} High($\geq3.0$)}                    & \cellcolor{mygreen!90}High($\geq30$)                                          & \cellcolor{myred!90}High($\geq2.0$)                                                     \\ \hhline{~----}
                            & GS-SLAM (2024)~\cite{yan2024gs}       & \cellcolor{myred!90}High($\geq3.0$)                                            & \cellcolor{mygreen!90}High($\geq30$)                                          & \cellcolor{myred!90}High($\geq0.5$)                                                     \\ \hhline{~----}
\multirow{-3}{*}{3DGS-SLAM} & Gaussian-SLAM (2023)~\cite{yugay2023gaussian} & \cellcolor{myred!90}High($\geq3.0$)                                          & \cellcolor{mygreen!90}High($\geq30$)                                          & \cellcolor{myred!90}High($\geq2.0$)                                                     \\ \hhline{-----}
                            & VoxFusion (2022)~\cite{yang2022vox}     & \cellcolor{myred!90}{\color[HTML]{000000} High($\geq3.0$)}                     & \cellcolor{myred!90}Low($\leq25$)                                        & \cellcolor{myred!90}High($\geq0.4$)                                                   \\ \hhline{~----}
\multirow{-2}{*}{Nerf-SLAM} & Nice-SLAM (2022)~\cite{zhu2022nice}     & \cellcolor{myred!90}High($\geq4.0$)                                             & \cellcolor{myred!90}Low($\leq25$)                                        & \cellcolor{mygreen!90}Low($\leq0.2$)                                                      \\ \hhline{-----}
                            & Droid-SLAM (2021)~\cite{teed2021droid}     & \cellcolor{mygreen!90}Low($\leq2.0$)                                             & \cellcolor{myred!90}Low($\leq20$)                                                    & \cellcolor{mygreen!90}Low($\leq0.2$)                                                      \\ \hhline{~----}
\multirow{-2}{*}{Trad-SLAM} & Orb-SLAM2 (2017)~\cite{mur2017orb}     & \cellcolor{mygreen!90}Low($\leq2.0$)                                            & \cellcolor{myred!90}Low($\leq20$)                                                    & \cellcolor{mygreen!90}Low($\leq0.2$)                                                      \\ \hhline{-----}
\end{tabular}%
}
\vspace{-8pt}
\end{table}

However, the exceptional scene representation capability of 3DGS-SLAM is achieved at the cost of deploying numerous Gaussians and multiple training iterations per frame. Specifically, conducting a representative 3DGS-SLAM algorithm, SplaTAM~\cite{keetha2024splatam}, for less than 600 frames on an Nvidia A100 GPU takes more than 20 minutes, demonstrating insufficient performance that limits its applications. For instance, autonomous robots can leverage a 3DGS-SLAM to support construction automation~\cite{yang2023redefining}, such as transporting building materials. The expectation is that the robots will complete scene modeling (training) within minutes and begin delivering promptly; otherwise, delays could lead to reduced productivity. Unfortunately, due to inadequate training performance, 3DGS-SLAM can still only learn environmental information slowly, highlighting the need for acceleration. While previous efforts, such as  GSCore~\cite{lee2024gscore} and Cicero~\cite{feng2024cicero}, successfully attains rapid inference speed for 3DGS or NeRF, they neglect to enhance the efficiency of the training process and lacks further investigation into streaming systems such as SLAM, thereby cannot achieve optimal performance.

In this paper, we begin by profiling 3DGS-SLAM and come to the recognition of three core performance bottlenecks. First, tracking utilizes numerous training iterations of 3DGS to estimate camera poses, leading to low throughput. Second, mapping overlooks the different contributions of Gaussians and leads to redundant computations. Specifically, a large portion of Gaussians do not contribute to scene representation but still consume substantial workloads. Lastly, we observe unbalanced workloads exist in both tracking and mapping, leading to severe hardware under-utilization.

\begin{figure}[!t]
\centering
\includegraphics[width=0.90\linewidth]{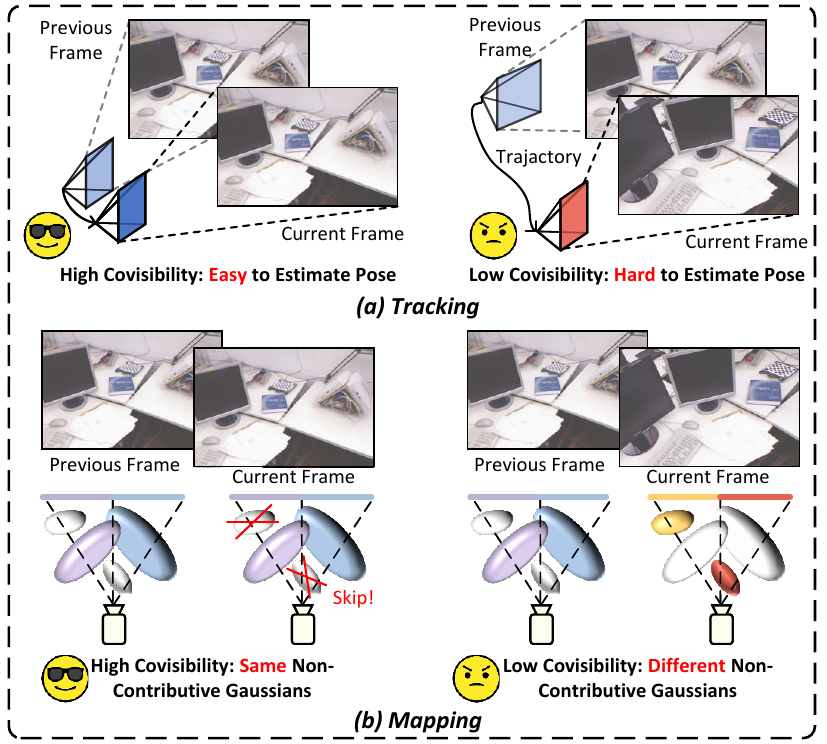}\vspace{-5pt}
\caption{Overview of AGS.}\vspace{-8pt}
\label{fig-Introduction}
\end{figure}

To address these limitations, we propose AGS, an algorithm-architecture co-design framework to accelerate 3DGS-SLAM based on one key insight that \textbf{\textit{SLAM systems process each incoming frame in a streaming manner, and consecutive frames exhibit high similarity that provides information for acceleration.}} From the software aspect, AGS proposes a two-level approach to optimize both the tracking and mapping tasks. 1) During tracking, we observe that the covisibility between frames (abbreviated as FC) indicates the magnitude of the robot's movement, leading to varying difficulty of pose estimation across frames. This motivates us to reduce redundant training iterations spent on frames with high covisibility. Specifically, as elaborated in Fig.~\ref{fig-Introduction} (a), a high FC implies minor movements where the pose of the current frame is easy to estimate from the previous frame, while a low FC indicates significant positional or perspective shifts, making pose estimation more challenging. Therefore, we adopt a lightweight algorithm to generate coarse estimations of pose for frames with high covisibility. For frames with low covisibility, the estimation is followed by fewer training iterations of 3DGS to fine-tune the pose. 2) During mapping, we observe that frame covisibility also indicates the similarity of Gaussian contribution information. As shown in Fig.~\ref{fig-Introduction} (b), a high FC implies that non-contributory Gaussians in the previous frame are more likely to remain non-contributory in the current frame. Thus, we can leverage the information from the previous frame to help identify and skip computations of non-contributory Gaussians for the current frame. In doing so, we designate key frames and non-key frames based on frame covisibility. For key frames, we perform full mapping and record the non-contributory Gaussians. For non-key frames, we utilize the recorded information to predict and skip redundant computations. 

From the hardware aspect, we introduce a customized
AGS architecture to support the AGS algorithm. To efficiently capture frame covisibility, we take a deep look into the System-on-Chip (SoC) in edge devices running SLAM and discover an hardware IP: the video CODEC, which is originally used to exploit covisibility across images for video compression~\cite{sullivan2012overview}. This observation inspired us to repurpose the intermediate results from the CODEC to define frame covisibility without adding significant hardware overhead. Moreover, we propose to disassemble the rendering pipeline of 3DGS and design dedicated hardware units to allow redistribution of the unbalanced workloads for tracking and mapping.

In summary, our key contributions are as follows: 

\begin{itemize}
\item  We present AGS, the first algorithm-architecture co-design framework to accelerate 3DGS-SLAM.

\item On the software level, we propose movement-adaptive tracking and Gaussian contribution-aware mapping to accelerate both tracking and mapping tasks based on different levels of frame covisibility.

\item On the hardware level, we first propose to leverage the intermediate results from CODEC as indicators to detect the frame covisibility in a lightweight manner. Then, we design specialized hardware units to ensure high-performance AGS algorithm execution and allow the redistribution of unbalanced workloads of 3DGS.

\end{itemize}

%% file: Background.tex
\section{Background}\label{sect:background}

\begin{figure*}[!t]
\centering
\includegraphics[width=0.85\linewidth]{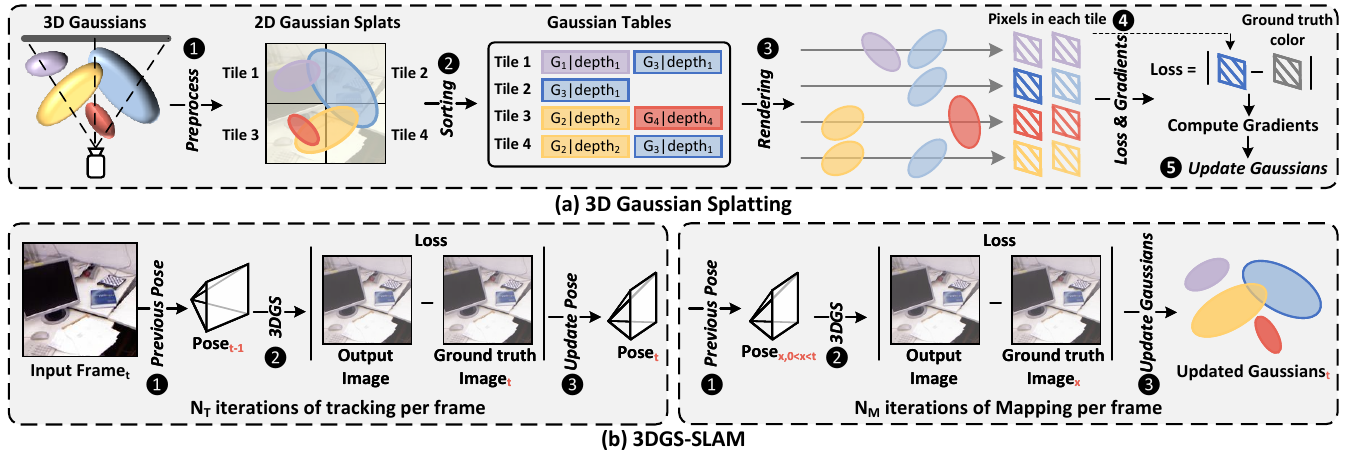}\vspace{-5pt}
\caption{Overview of 3D Gaussian Splatting and 3DGS-SLAM.}\vspace{-5pt}
\label{fig-3DGS}
\end{figure*}

\subsection{Preliminaries of 3DGS}\label{ssect:3DGS}
3DGS represents a breakthrough in 3D reconstruction tasks and has been applied across a spectrum of domains, such as autonomous driving~\cite{cao2024lightning, guo2023streetsurf, nguyen2022snerf, yang2024unipad}, embodied intelligence~\cite{adamkiewicz2022vision, byravan2023nerf2real, chen2024catnips, hu2023nerf, maggio2023loc, wu2023mapnerf}, and augmented reality (AR)~\cite{deng2022fov, li2024magic}. It explicitly represents scenes using anisotropic Gaussian ellipsoids, with each Gaussian characterized by attributes of position, shape, color, etc., collectively referred to as Gaussian features. Each iteration of 3DGS training contains the following steps:

Given a camera viewpoint, \textbf{step $\hquan{1}$} performs \textit{preprocessing} to project visible Gaussians from 3D space to the 2D imaging plane according to the camera intrinsics. Based on the 2D Gaussians, we can determine the corresponding tiles ($n \times n$ pixels) associated with each Gaussian by an intersection test. As illustrated in Fig.~\ref{fig-3DGS} (a), the orange Gaussian intersects with tile 3 and tile 4 and thus will be included in the rendering process of all pixels in both tiles. Subsequently, \textbf{step $\hquan{2}$} performs \textit{depth sorting} of Gaussians to generate a Gaussian table for each tile to record the indexes (IDs) and depth sequence of its related Gaussians. With the guidance of Gaussian tables, \textbf{step $\hquan{3}$} conducts \textit{rendering} with Gaussian features to compute the color of pixels ($C$) in two stages. The first stage computes the occlusion factor $\alpha$ by Eqn. (1).
 
\vspace{-8pt}
\begin{align}\label{eqn-1}
    \alpha_i = exp(-{\frac{1}{2}}(x-\mu_{i}){\Sigma}_{i}^{-1}(x-\mu_{i}))
\end{align}
\vspace{-8pt}

\noindent where $\mu_i$ and $\Sigma_i$ are Gaussian features describing positions and shapes, and $x$ is the coordinates of the pixel. The second stage uses alpha-blending technique to compute the transmittance $T_i = \prod_{j=1}^{i-1} {(1-\alpha_j)}$ and traverses the depth-ordered Gaussians from front to back to accumulate the pixel color with $\alpha_i$, $T_i$, and the color of each Gaussian $c_i$ by Eqn. (2). 

\vspace{-8pt}
\begin{align}\label{eqn-1}
    {C} &= \sum_{i=1}^nT_i\alpha_ic_i
\end{align}
\vspace{-8pt}

\noindent where $n$ is the number of Gaussians associated with the pixel. Moreover, $\alpha$ serves as a key indicator of a Gaussian's contribution to a particular pixel, with larger $\alpha$ values signifying a greater influence on the pixel's color. Finally, if the value of $T$ becomes smaller than a threshold ($10^{-4}$), the rendering for the pixel terminates (early termination). After rendering, \textbf{step $\hquan{4}$} executes \textit{gradient computation} that computes the gradients of Gaussians based on the loss generated from the rendered pixels. Last, \textbf{step $\hquan{5}$} \textit{updates Gaussians} with their gradients to finish one training iteration.

\subsection{3DGS-SLAM}

SLAM typically consists of two main tasks, tracking and mapping. Tracking estimates the robot's current position, while mapping incrementally reconstructs a 3D representation of the scene. As the robot navigates in an unknown environment, it consecutively captures frames of images to serve as input for both tracking and mapping. After sufficient frames are processed and the scene is well-reconstructed, the online training process terminates. In subsequent operations, the robot can obtain geometry and photometric properties (e.g. depth and color) through an inference process. These properties can then be employed to facilitate the robot's path planning and motion control.

Integrating SLAM with 3DGS enables high-fidelity scene reconstruction. For each input frame, the tracking and mapping of 3DGS-SLAM proceed as follows: As illustrated in Fig.~\ref{fig-3DGS} (b), to estimate the camera pose of the current frame ($Pose_t$), tracking first performs \textbf{step $\hquan{1}\rightarrow$ step $\hquan{4}$} in \ref{ssect:3DGS} with the pose from the previous frame ($Pose_{t-1}$) to obtain the gradients of Gaussians. Instead of updating Gaussians, we further derive gradients of pose from gradients of Gaussians to update the pose for $N_{T}$ iterations while keeping Gaussians unchanged, and the result serves as the current pose ($Pose_t$). Subsequently, mapping fixes the camera pose and completes \textbf{step $\hquan{1}\rightarrow$ step $\hquan{5}$} for $N_{M}$ iterations to update Gaussians. Notably, mapping utilizes not only the current pose ($Pose_t$) for training, but also other poses ($Pose_x, 0 \textless x \textless t$) and images from previous frames. In summary, each frame in 3DGS-SLAM requires multiple iterations of 3DGS. Due to the vast number of Gaussians and numerous training iterations, both tracking and mapping exhibit insufficient throughput.

\begin{figure}[!t]
\centering
\includegraphics[width=0.85\linewidth]{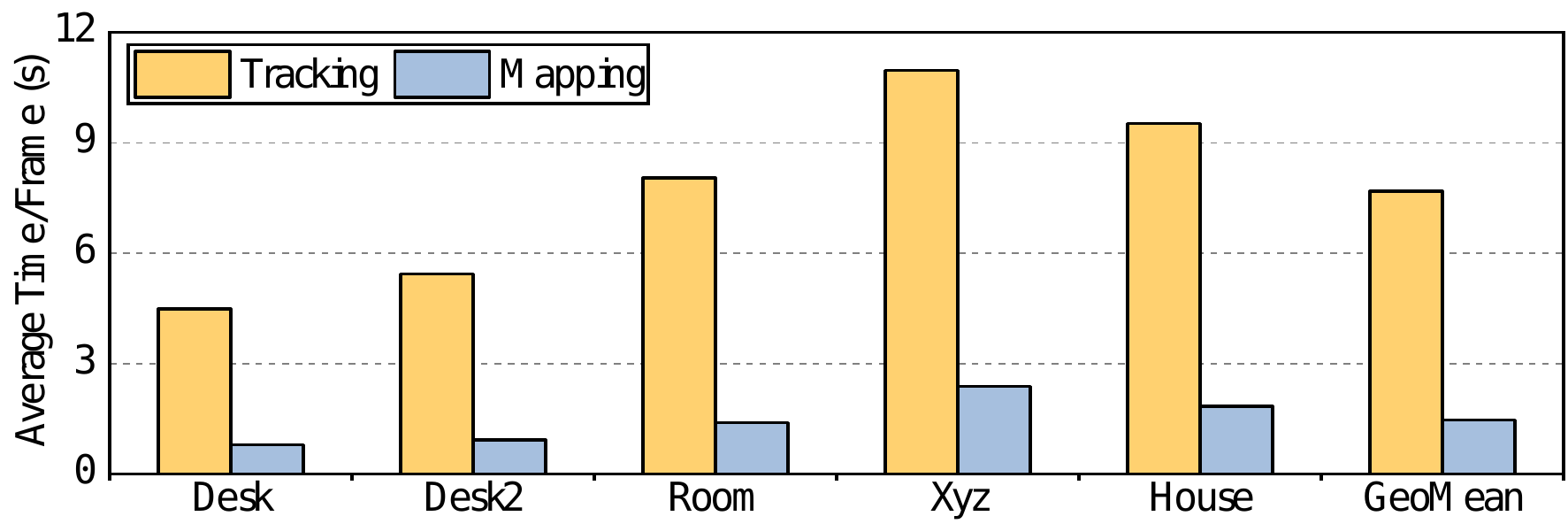}\vspace{-5pt}
\caption{Execution time of tracking and mapping.}\vspace{-5pt}
\label{fig-Time}
\end{figure}

\begin{figure}[!t]
\centering
\includegraphics[width=0.85\linewidth]{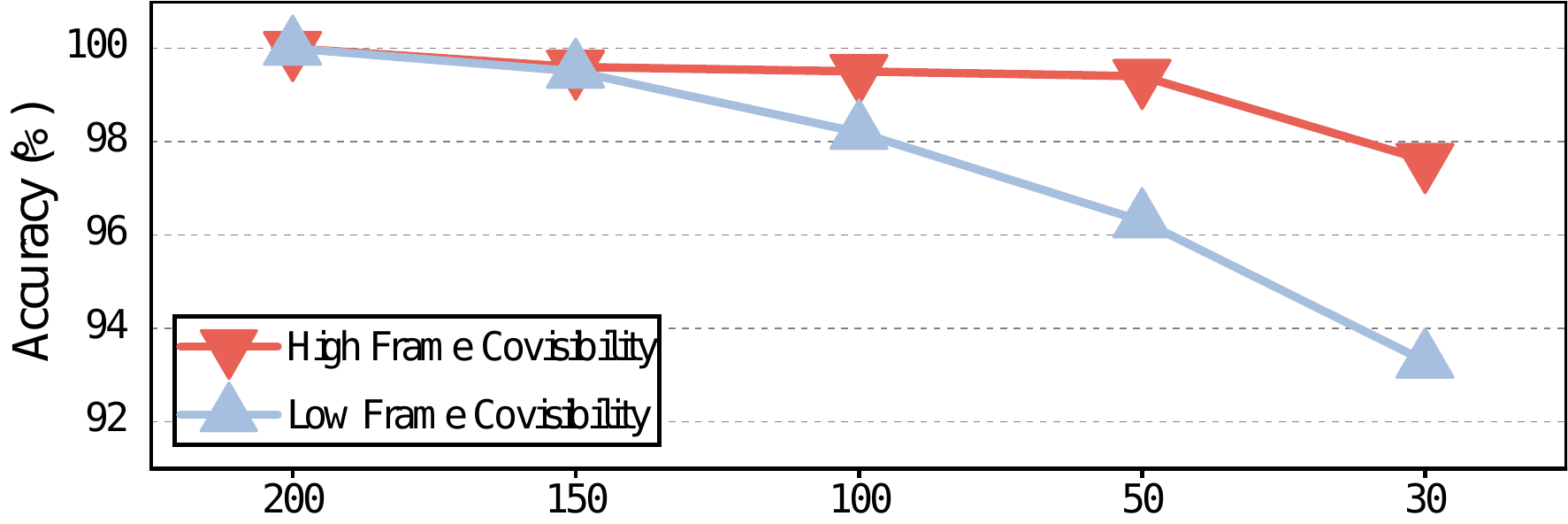}\vspace{-5pt}
\caption{Accuracy loss with reduced training iterations.}\vspace{-5pt}
\label{fig-Motivation_t}
\end{figure}

\subsection{Video CODEC}\label{ssect:codec}

CODEC units are commonly deployed on digital circuits and are mainly used for video compression. To maximize the compression ratio while maintaining video quality, motion estimation (ME) is a core algorithm to search for the most similar macro-block (MB) between consecutive frames.

The ME algorithm divides the current frame and the previous frame into multiple MBs (e.g. $8\times8$ pixels) for matching. For each MB in the current frame, the algorithm searches in the previous frame by calculating the Sum of Absolute Differences (SAD) between the two MBs. Intuitively, a smaller SAD value indicates greater similarity between two MBs, while a larger value indicates less. Therefore, the MB with the smallest SAD value is selected as the matching MB for the current one to describe how MBs move between consecutive frames. In this paper, instead of acquiring the movement of MBs, we focus on leveraging the SAD values computed by the video codec. Since these values reflect the content similarity between frames, we can extract them as a quantitative metric for evaluating frame covisibility.

%% file: Motivation.tex
\section{Motivation}\label{sect:motivation}

To analyze the performance bottlenecks of 3DGS-SLAM, we run the implementation of a representative framework, SplaTAM, on an A100 GPU with the commonly used TUM-RGBD dataset~\cite{sturm12iros}. After careful investigation, we first break down the inefficiency of 3DGS-SLAM into the following challenges and observations.

\textbf{Challenge 1:} \textbf{\textit{Tracking requires numerous training iterations to converge.}} As depicted in Fig.~\ref{fig-Time}, tracking consumes $83\%$ of the time and yields a lower frame rate than mapping. This issue arises because tracking typically needs 200 training iterations to achieve ideal accuracy, while mapping only requires 30. This calls for a faster convergence method to reduce the training iterations spent on tracking.

\begin{figure}[!t]
\centering
\includegraphics[width=0.85\linewidth]{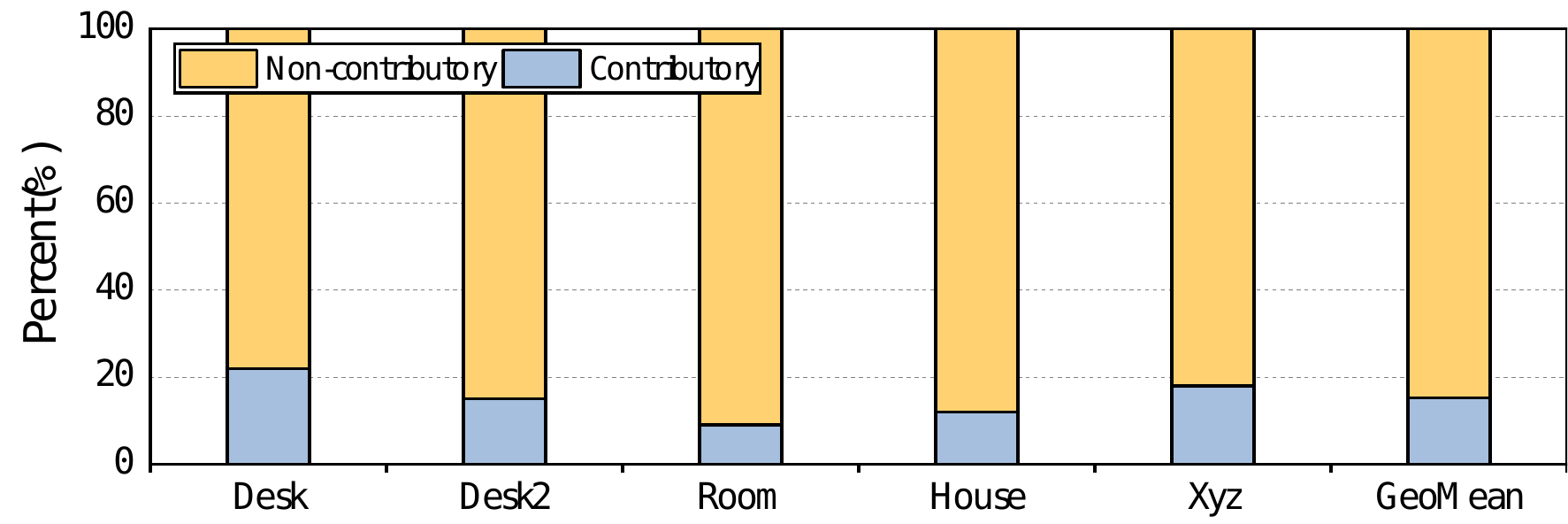}\vspace{-5pt}
\caption{Gaussians with different contributions to pixels during rendering.}\vspace{-5pt}
\label{fig-Unused}
\end{figure}

\begin{figure}[!t]
\centering
\includegraphics[width=0.85\linewidth]{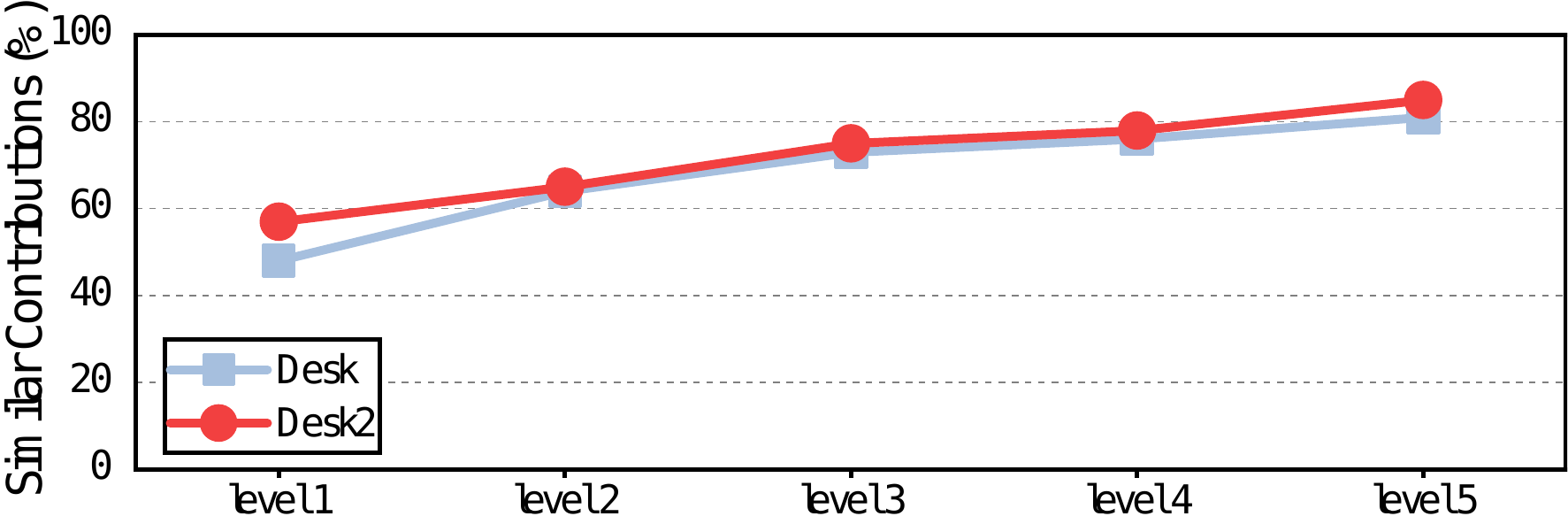}\vspace{-5pt}
\caption{Contribution similarity of Gaussians between frames with different levels of covisibility. }\vspace{-5pt}
\label{fig-Motivation_m}
\end{figure}

\textbf{Observation 1:} \textbf{\textit{Varying frame covisibility (FC) leads to excessive training iterations.}} We observe that variations in frame covisibility result in differing sensitivities to reductions in training iterations, as depicted in Fig.~\ref{fig-Motivation_t}. Specifically, we gradually reduce the training iterations of the frames with high/low FC. The results show that frames with low covisibility are more sensitive, experiencing a greater accuracy loss of $6.7\%$. This indicates that numerous training iterations spent for frames with high covisibility are unnecessary, leaving room for reducing the training iterations.

\begin{figure*}[!t]
\centering
\includegraphics[width=0.85\linewidth]{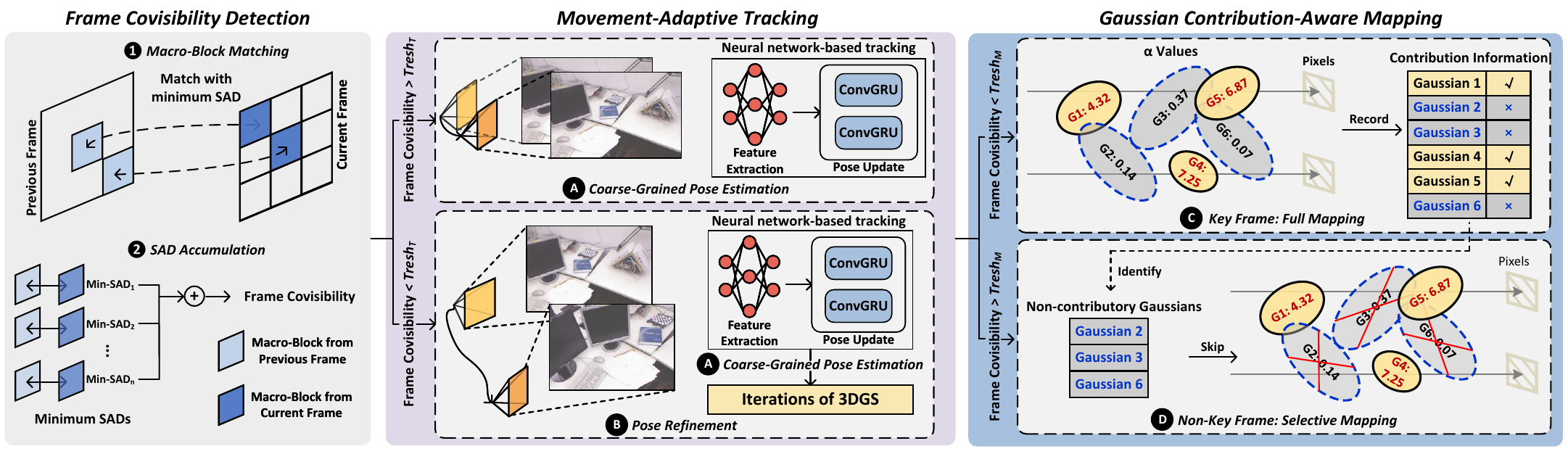}\vspace{-0pt}
\caption{Overview of the AGS Algorithm.}\vspace{-0pt}
\label{fig-Algorithm_overview}
\end{figure*}

\textbf{Challenge 2:} \textbf{\textit{Redundant computation workloads on non-contributory Gaussians.}} For each training iteration, 3DGS involves a vast number of Gaussians, many of which contribute none to the color of pixels. As shown in Fig.~\ref{fig-Unused}, among all Gaussians assigned to one Gaussian table, 85.1\% of the Gaussians have no impact on the pixel color. This motivates us to seek methods to efficiently identify and skip the computation of these non-contributory Gaussians. 

\textbf{Observation 2:} \textbf{\textit{The opportunity of skipping the computations of non-contributory Gaussians.}} Since a large portion of Gaussians has no impact on the pixel colors, a viable solution is to identify and skip the computation of those non-contributory Gaussians in runtime with minimum overhead. Luckily, we observe that the contributions of Gaussians are similar in frames with high covisibility. 

As demonstrated in Fig.~\ref{fig-Motivation_m}, we categorize FC into five levels, where a higher level indicates higher covisibility with the previous frame. Results show that in frames with level-5 FC, more than $80\%$ of the non-contributory Gaussians from the previous frame remain non-contributory in the current frame. This provides us with the opportunity to predict and skip the computation of non-contributory Gaussians with the guidance of information from previous frames.

\textbf{Challenge 3:} \textbf{\textit{Unbalanced workloads in both tracking and mapping.}} Due to the early-termination effect discussed in section~\ref{ssect:3DGS}, some pixels complete rendering in advance of others, leaving the computation units responsible for those pixels idle while waiting for the remaining ones to finish. Moreover, skipping the predicted non-contributory Gaussians aggravates the unbalanced workloads among these units, leading to significant underutilization of hardware resources.

\textbf{Observation 3:} \textbf{\textit{Disassembling the rendering process enables the redistribution of unbalanced workloads.}} To address challenge 3, a viable solution is to distribute the workload of busy units to idle units. To overcome the data dependency caused by the recurrence calculation of alpha-blending in Eqn. (2), we propose to disassemble the rendering process of each Gaussian into two stages. We observe that the first stage of alpha computation operates independently of previous operations. This enables the redistribution of its workload from busy units to idle ones by fine-grained scheduling, thereby alleviating the unbalanced workloads.

In all, we aim to leverage frame covisibility for accelerating both tracking and mapping while mitigating the unbalanced workloads during training, leaving three more problems to be solved: 1) Identification of frame covisibility during runtime. 2) Utilizing frame covisibility to accelerate tracking and mapping. 3) Breaking data dependencies of alpha-blending to enable workload redistribution between pixels. 

%% file: Algorithm.tex
\section{Algorithm}\label{sect:algorithm}

\subsection{Algorithm Overview}

The key insight of the AGS algorithm is to leverage frame covisibility to accelerate both tracking and mapping. In this section, we propose CODEC-based frame covisibility detection, movement-adaptive tracking, and Gaussian contribution-aware mapping to overcome the performance bottlenecks of 3DGS-SLAM, as depicted in Fig.~\ref{fig-Algorithm_overview}. First, we propose to identify and quantify frame covisibility with the CODEC. During the ME algorithm detailed in~\ref{ssect:codec}, the CODEC computes SAD values and uses the minimum SAD to identify the matching MBs across consecutive frames. As the SAD values quantify the difference between the two MBs, we save the minimum SADs ($SAD_{min}^i$) for all MBs and accumulate them by $\sum_{i}SAD_{min}^i$, with a greater accumulation result indicating less frame covisibility and vice versa. Subsequently, the frame covisibility is utilized to instruct both tracking and mapping to perform movement-adaptive tracking and Gaussian contribution-aware mapping.

\subsection{Movement-Adaptive Tracking}

For tracking, frame covisibility reflects the magnitude of the robot's movement between two consecutive frames. With each incoming frame, we first conduct a \textit{coarse-grained pose estimation} ($\hquan{A}$ in Fig.~\ref{fig-Algorithm_overview}) that employs a lightweight algorithm inspired by neural network-based tracking approaches to generate a coarse-grained estimation of the camera pose, and selectively execute a \textit{fine-grained pose refinement} ($\hquan{B}$) depending on the covisibility between the current frame and the previous frame. 

For frames with covisibility higher than a pre-set threshold $Thresh_{T}$, implying minor changes in the robot’s position and orientation, the coarse-grained pose estimation is sufficient for the subsequent mapping task. For frames with covisibility lower than $Thresh_{T}$, which likely involve more movement of the robot, we follow up the estimation with $Iter_{T}$ training iterations of 3DGS as a fine-grained refinement. Note that $Iter_{T}$ is significantly fewer than the baseline training iterations, thus minimizing the overall convergence time. The coarse-grained pose estimation builds on the backbone of Droid-SLAM~\cite{teed2021droid}, which delivers much faster convergence speed than training 3DGS. It first extracts frame features by a convolutional neural network and then utilizes GRUs to update the current pose, which is more hardware-friendly compared to training 3DGS due to the simpler control logic and high computation throughput of matrix multiplication and convolution operations.

With movement-adaptive tracking reducing the training iterations for tracking, mapping now accounts for a larger proportion of the execution time, highlighting the need for optimization. Therefore, we propose Gaussian contribution-aware mapping based on observation 2 in section~\ref{sect:motivation}.

\begin{figure}[!t]
\centering
\includegraphics[width=0.90\linewidth]{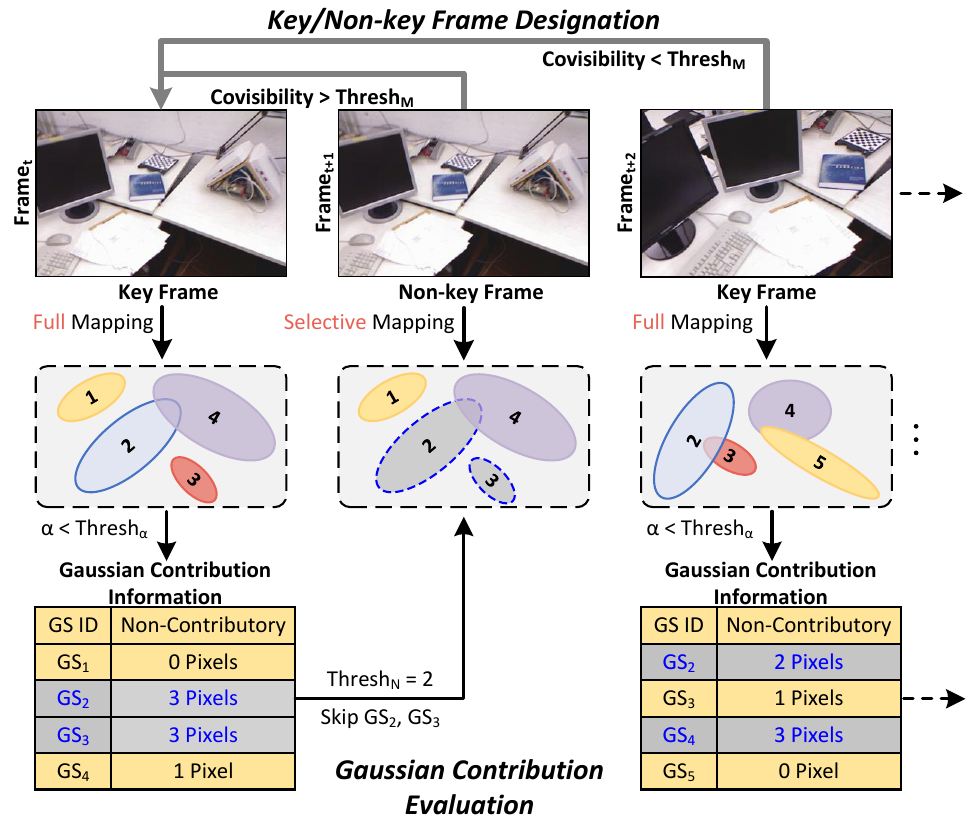}\vspace{-0pt}
\caption{Details of Gaussian contribution-aware mapping.}\vspace{-5pt}
\label{fig-Mapping}
\end{figure}

\subsection{Gaussian Contribution-Aware Mapping}

For mapping, higher frame covisibility implies a greater similarity in the contribution of Gaussians. Therefore, we propose Gaussian contribution-aware mapping to utilize the Gaussian contribution information from previous frames as predictions, enabling us to identify and skip the computation of non-contributory Gaussians of the current frame. In doing so, we categorize all frames into key frames and non-key frames: For key frames, we run \textit{full mapping} ($\hquan{C}$ in Fig.~\ref{fig-Algorithm_overview}) that performs the baseline 3DGS algorithm to record the contribution information. For non-key frames, we reuse the recorded information to execute \textit{selective mapping} ($\hquan{D}$) that skips the non-contributory Gaussians for acceleration. The remaining problems are how to designate key/non-key frames, and how to evaluate the contribution of Gaussians.

\textbf{Key/non-key frame designation.} First, we designate key frames and non-key frames based on the principle that non-key frames exhibit high frame covisibility with key frames, while key frames maintain low frame covisibility with one another. Therefore, we obtain the frame covisibility of the current frame with a previous key frame from the CODEC and then compare it with a pre-set threshold, $Thresh_{M}$. As illustrated in Fig.~\ref{fig-Mapping}, assume $frame_{t}$ is a previous key frame that has executed full mapping and recorded the Gaussian contribution information. For an incoming frame ($frame_{t+1}$), we obtain its frame covisibility with $frame_{t}$ and designate it as a non-key frame given that the frame covisibility is higher than $Thresh_{M}$. As a result, $frame_{t+1}$ executes selective mapping to skip the non-contributory Gaussians predicted by $frame_{t}$. For the next frame ($frame_{t+2}$) with low covisibility, we designate it as a new key frame, execute full mapping, and update the Gaussian contribution information accordingly.

\textbf{Gaussian contribution evaluation.} The evaluation of Gaussian contribution stems from two aspects: the $\alpha$ value, and the number of pixels to which the Gaussian makes negligible contributions. First, we choose the $\alpha$ value as an indicator of Gaussian contribution to a pixel. If a Gaussian’s $\alpha$ value is below a certain threshold $Thresh_{\alpha}$, it’s marked as non-contributory for the corresponding pixel. Next, since each Gaussian is related to multiple pixels, we choose the non-contributory number of a Gaussian as an indicator of contribution to the non-key frames. If a Gaussian has negligible contribution to more than $Thresh_{N}$ pixels, it's marked as non-contributory for the following non-key frames. As shown in Fig.~\ref{fig-Mapping}, $GS_{2}$ and $GS_{3}$ (colored in gray) have $\alpha$ values lower than $Thresh_{\alpha}$ for three pixels, so both their non-contributory numbers are recorded as 3 in the Gaussian contribution information. To further instruct the non-key frames with recorded information, we compare the non-contributory numbers with $Thresh_{N}$. Since both $GS_{2}$ and $GS_{3}$ are non-contributory to $\geq2$ pixels, we exclude them in the following non-key frames for selective mapping.

\begin{figure}[!t]
\centering
\includegraphics[width=0.90\linewidth]{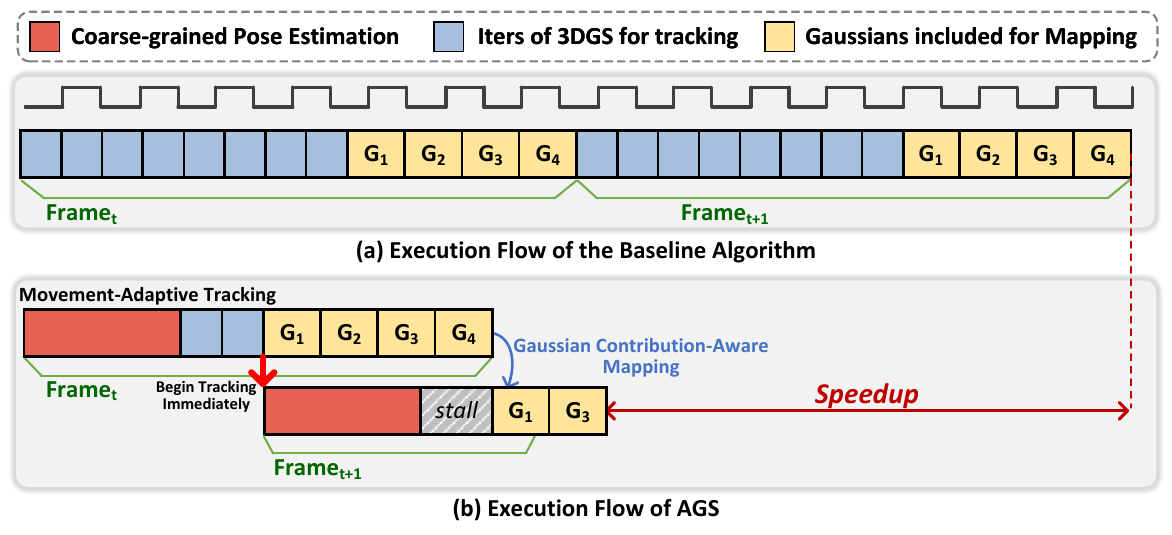}\vspace{-0pt}
\caption{A walk-through example of AGS algorithm.}\vspace{-5pt}
\label{fig-Walkthrough}
\end{figure}

\begin{figure*}[!t]
\centering
\includegraphics[width=0.90\linewidth]{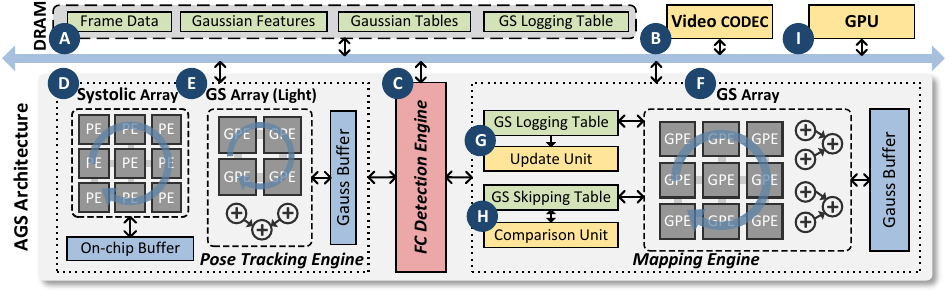}\vspace{-0pt}
\caption{Overview of AGS Architecture.}\vspace{-5pt}
\label{fig-Architecture_overview}
\end{figure*}

Additionally, we observe that the pre-set hyperparameters show an impact on performance and accuracy, including $Thresh_{T}$, $Iter_{T}$, $Thresh_{M}$, $Thresh_{\alpha}$, and $Thresh_{N}$. All hyper-parameters help to reduce redundant workloads of 3DGS-SLAM while inducing accuracy loss. To strike a balance between performance and accuracy, we set $Thresh_{T}$ as $90\%$, $Thresh_{\alpha}$ as $\frac{1}{255}$, and determine suitable values for the other parameters with further discussion in Section~\ref{ssect:exploration}.

\subsection{AGS: A Walk-Through Example}

With the adoption of coarse-grained pose estimation in the movement-adaptive tracking algorithm, the tracking of SLAM no longer relies on the Gaussians updated by mapping, enabling overlapped execution of the two tasks. To demonstrate the advantage of movement-adaptive tracking, Gaussian contribution-aware mapping, and a pipelined execution flow, we provide a walk-through example of the AGS algorithm.

Fig.~\ref{fig-Walkthrough} (a) illustrates the execution flow of the baseline 3DGS-SLAM system. Each new frame undergoes tracking with numerous training iterations of 3DGS to produce the pose, which then triggers mapping for updating Gaussians. The updated Gaussians are then used for the tracking task of the subsequent frame, demanding a serialized process. 

Fig.~\ref{fig-Walkthrough} (b) demonstrates the execution flow of the proposed AGS algorithm based on movement-adaptive tracking and Gaussian contribution-aware mapping. For $frame_{t}$, the coarse-grained pose estimation, colored in red, efficiently generates an estimated pose. Next, given that $frame_{t}$ has a low covisibility with its previous frame, a fine-grained pose refinement is required. After obtaining the refined pose, the mapping of $frame_{t}$ is executed to run full mapping, and the contribution information of Gaussians is recorded. In the meantime, since the coarse-grained pose estimation of $frame_{t+1}$ does not rely on the Gaussians updated from $frame_{t}$, it is initiated immediately after the fine-grained pose refinement of $frame_{t}$ is completed. Given a high covisibility for $frame_{t+1}$ with $frame_{t}$, we eliminate the need for fine-grained pose refinement and conduct selective mapping based on the contribution information saved from $frame_{t}$. As a result, numerous iterations of tracking are reduced and a significant proportion of Gaussians are excluded for mapping, thus boosting the performance of the SLAM system.

%% file: Architecture.tex
\section{Architecture}\label{sect:architecture}

\subsection{Architecture Overview}

This section presents the AGS architecture. As illustrated in Fig.~\ref{fig-Architecture_overview}, it comprises an FC detection engine, a pose tracking engine, and a mapping engine. Since the SAD values are naturally generated by the video CODEC, it requires little overhead for the FC detection engine to extract these values through main memory. It then accumulates the SADs to generate frame covisibility and compares it with predefined thresholds to instruct the following executions. For movement-adaptive tracking, the pose tracking engine utilizes a systolic array and a lightweight GS array to perform coarse-grained pose estimation and fine-grained pose refinement, respectively. To support Gaussian contribution-aware mapping, the mapping engine is equipped with a full-scale GS array, a GS logging table, and an update unit to perform full mapping and capture Gaussian contribution information for key frames. For non-key frames, we design a GS skipping table and a comparison unit to identify the non-contributory Gaussians and support selective mapping. 

The data flow of AGS for processing an incoming frame can be divided into three steps: 1) Computing frame covisibility through the cooperation of the CODEC and the FC detection engine ($\hquan{A}\rightarrow\hquan{B}\rightarrow\hquan{C}$). Specifically, the CODEC conducts the ME algorithm and stores the resultant SAD values in DRAM, which will later be accessed by the FC detection engine to compute frame covisibility. 2) The second step executes movement-adaptive tracking with the pose tracking engine ($\hquan{D}\rightarrow\hquan{E}$). 3) The third step involves the mapping engine to conduct full mapping for key frames ($\hquan{F}\rightarrow\hquan{G}\rightarrow\hquan{A}$) or selective mapping for non-key frames ($\hquan{A}\rightarrow\hquan{H}\rightarrow\hquan{F}$). Due to independent hardware resource allocation, while mapping is in progress, we can trigger the FC detection engine and the pose tracking engine to execute the next frame's FC detection and pose estimation, enabling the overlap of tracking and mapping.

\subsection{Mapping Engine}\label{ssect:map}

The mapping engine supports full mapping for key frames and selective mapping for non-key frames. However, both recording and utilizing the Gaussian contribution information pose challenges due to the vast number of Gaussians. Specifically, the Gaussian contribution information exceeds the on-chip memory capacity and must be kept off-chip. To ensure memory consistency, recording or retrieving them causes frequent DRAM access, which occupies heavy bandwidth and hinders the overall performance. Therefore, we propose a speculation method to identify the `hot' Gaussians that will be accessed more frequently during a period. In this way, we reserve their associated contribution information on-chip for a long time to avoid repeated access. 

\begin{figure}[!t]
\centering
\includegraphics[width=0.9\linewidth]{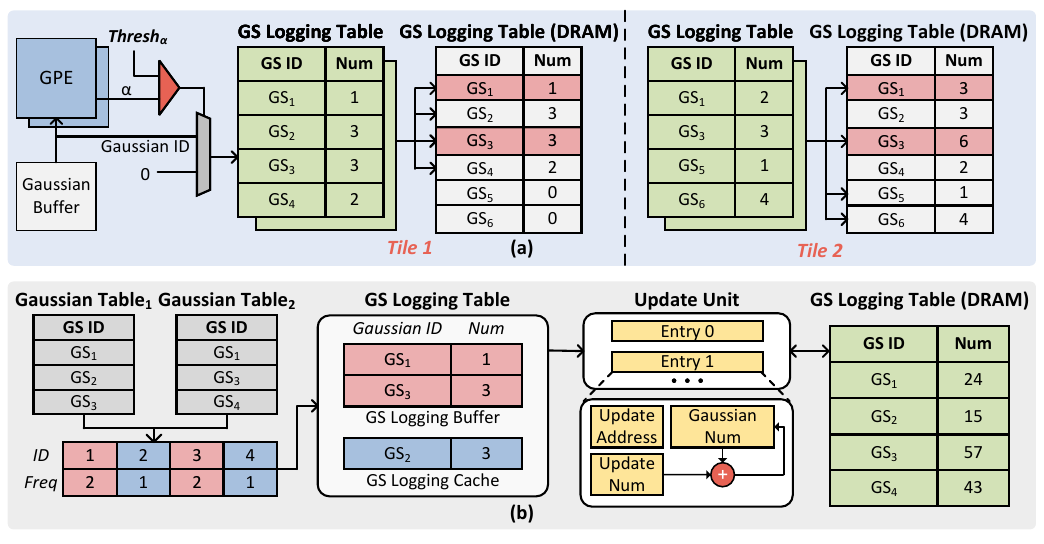}\vspace{-10pt}
\caption{The GS logging table and the update unit.}\vspace{-10pt}
\label{fig-Update}
\end{figure}

\textbf{GS logging table and the update unit.} The GS logging table and the update unit are activated only for key frames to capture and update the Gaussian contribution information to off-chip memory during full mapping. As depicted in Fig.~\ref{fig-Update} (a), by extracting the $\alpha$ values computed by GPEs and comparing them with $Thresh_{\alpha}$, we can increment the non-contributory number of a Gaussian by its ID. As the 3DGS algorithm executes in a tile-by-tile manner, a naive solution is to accommodate the related Gaussian contribution information with a limited on-chip buffer and sequentially update them to off-chip memory immediately after a tile is finished, which leads to repeated DRAM access for the same Gaussians across multiple tiles, such as $GS_{1}$ and $GS_{3}$.

Fig.~\ref{fig-Update} (b) demonstrates how the proposed GS logging table collaborates with the update unit to address the above issue. Before rendering, the GS logging table fetches Gaussian tables of multiple tiles to evaluate the accessed frequency of Gaussians. For instance, $GS_{1}$ and $GS_{3}$ appear in both tables, indicating they are `hot' Gaussians whose information will be repeatedly updated to DRAM. Accordingly, we equip the GS logging table with a GS logging buffer to preserve the contribution information of the `hot' Gaussians, while information of the `cold' Gaussians with low accessed frequency is handled by a GS logging cache. For `cold' Gaussians, after incrementing the non-contributory numbers of Gaussians for one tile, the update unit fetches the numbers of the corresponding Gaussians from DRAM, adds them with the numbers from the GS logging cache, and writes the accumulated results back. On the contrary, for `hot' Gaussians, the non-contributory numbers in the GS logging buffer are preserved and finally updated to DRAM until all the evaluated tiles are finished, reducing a considerable amount of DRAM access.

\begin{figure}[!t]
\centering
\includegraphics[width=0.9\linewidth]{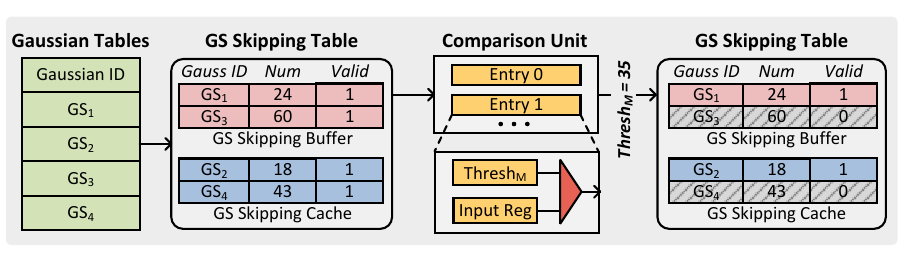}\vspace{-10pt}
\caption{The GS skipping table and the comparison unit.}\vspace{-10pt}
\label{fig-Comp}
\end{figure}

\textbf{GS skipping table and the comparison unit.} The GS skipping table and the comparison unit are utilized to perform selective mapping for non-key frames. As shown in Fig.~\ref{fig-Comp}, the Gaussian skipping table initially retrieves Gaussian contribution information from DRAM. Note that the GS skipping table is also equipped with a GS skipping buffer and a GS skipping cache to reduce reading requests of Gaussians from off-chip memory. To identify and eliminate the non-contributory Gaussians, each entry of the Gaussian skipping table is assigned a Gaussian ID, its non-contributory number, and a valid flag originally set to 1. The comparison unit evaluates the non-contributory number of each Gaussian against $Thresh_{M}$. For Gaussians whose non-contributory number is larger than $Thresh_{M}$, the comparison unit sets the corresponding valid flags to 0 to make the GS array skip the computations on them. Afterward, the GS array follows the instruction of the GS skipping table instead of the original Gaussian tables to only fetch related Gaussians that are valid for selective mapping. For instance, with $Thresh_{M}$ set to 35, $GS_{3}$ and $GS_{4}$ in gray boxes are predicted as non-contributory and will be skipped in the current frame. 

\subsection{Pose Tracking Engine}\label{ssect:pose}

The pose tracking engine performs the coarse-grained pose estimation and selectively executes pose refinement based on frame covisibility. First, it consists of a set of $32\times32$ systolic arrays and an on-chip buffer to support coarse-grained pose estimation. Subsequently, the control signal from the FC detection engine determines whether to activate the lightweight GS array to perform fine-grained pose refinement. The lightweight GS array consists of several $4\times4$ GPEs to perform rendering and gradient computation of $4\times4$ pixels. 

During rendering, each GPE calculates pixel color based on Gaussian features, involving two main stages: $\alpha$ computation and color rendering. First, the GPE calculates the $\alpha$ value by Eqn. (1). Next, $\alpha$ is utilized to iteratively update the transmittance ($T$) and accumulate the color of pixels by Eqn. (2). For gradient computation, the GPE calculates Gaussian gradients based on pixel loss and Gaussian features. To merge gradients of the same Gaussian generated by multiple pixels, each GPE array is equipped with an adder tree to accumulate gradients efficiently.

\begin{figure}[!t]
\centering
\includegraphics[width=1\linewidth]{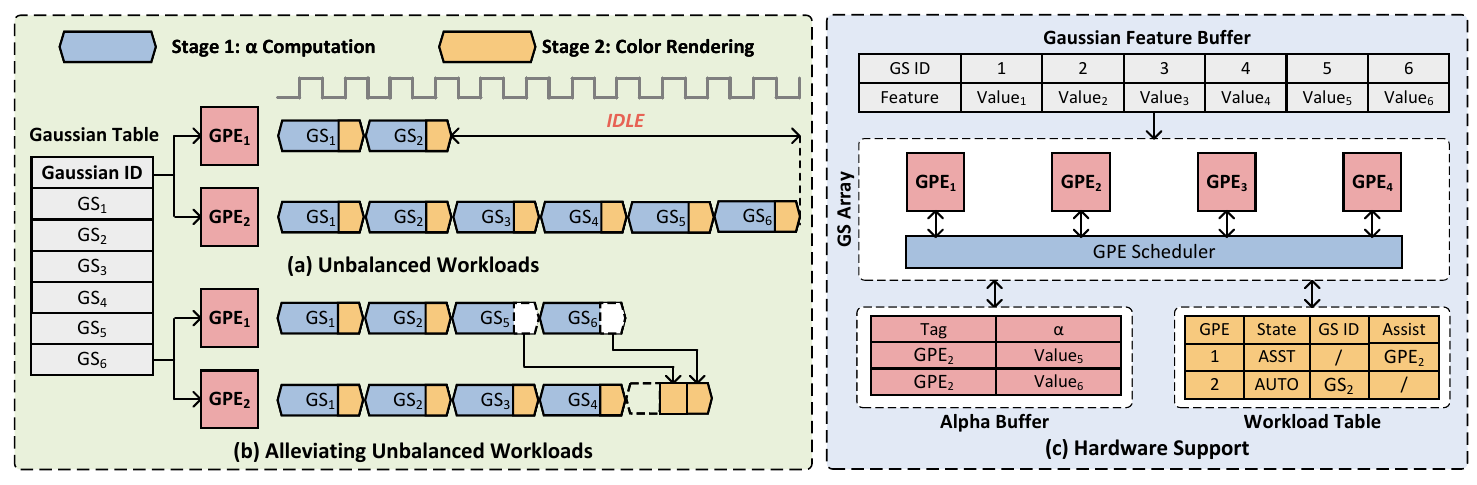}\vspace{-10pt}
\caption{Alleviating unbalanced workloads between GPEs.}\vspace{-10pt}
\label{fig-Balance}
\end{figure}

\subsection{GPE Scheduler Design}

As discussed in~\ref{ssect:3DGS}, early termination happens when the transmittance $T$ falls below a threshold, resulting in a workload unbalance problem. Even worse, the proposed Gaussian contribution-aware mapping method exacerbates the unbalanced workloads due to the unpredictable number of non-contributory Gaussians. As depicted in Fig.~\ref{fig-Balance} (a), the rendering of $GPE_{1}$ completes after processing two Gaussians, while $GPE_{2}$ continues until it has computed six Gaussians. As a result, $GPE_{1}$ is idle for $66\%$ of the total execution time. 

To alleviate the unbalanced workloads and maximize hardware utilization, we propose a GPE scheduler, a workload table, and an alpha buffer to redistribute the workload of unfinished GPEs to idle GPEs with a dynamic scheduling scheme. The proposal is built on an observation that during rendering, while the color rendering (the second stage) involves the recursive calculation of $T$ that incurs data dependencies, the $\alpha$ computation (the first stage) is independent of previous results and accounts for the majority of the rendering time. Therefore, we propose to disassemble the two stages of each Gaussian and distribute the first stage's workload to idle GPEs. In doing so, we define two working states for GPEs: the autonomous state and the assistant state. To assist others, a $GPE_{m}$ in the assistant state executes the first stage for another $GPE_{n}$ in advance and stores the $\alpha$ results in the alpha buffer with a tag of $n$. Correspondingly, $GPE_{n}$ in the autonomous state, checks the alpha buffer periodically and obtains the pre-computed $\alpha$ values by hitting the tag. 

As depicted in Fig.~\ref{fig-Balance} (b) and (c), after $GPE_{1}$ finishes, the GPE scheduler traverses through the workload table and assigns $GPE_{2}$ for it to assist. Subsequently, $GPE_{1}$ fetches the Gaussian features of $GS_{5}$ and $GS_{6}$ accordingly to perform $\alpha$ computation, and stores the results in the alpha buffer with a tag corresponding to $GPE_{2}$. By the time $GPE_{2}$ completes $GS_{4}$, it checks the alpha buffer and successfully hits the tag so as to obtain the $\alpha$ values pre-computed by $GPE_{1}$. Finally, $GPE_{2}$ completes rendering for $GS_{5}$ and $GS_{6}$ by executing their second stages.

%% file: Experiment.tex
\section{Evaluation}\label{sect:eval}

\begin{table}[!t]
\centering
\renewcommand\arraystretch{1.25}
\caption{Tracking Accuracy (ATE RMSE$\downarrow$ [cm]).}
\label{tab_Ate}
\setlength{\arrayrulewidth}{0.85pt}
\resizebox{0.9\columnwidth}{!}{%
\begin{tabular}{c|cccccc}
\hline
         & Desk & Desk2 & Room  & Xyz  & House & GeoMean \\ \hline
\begin{tabular}[c]{@{}c@{}}SplatAM (3DGS)\end{tabular}  & 3.30  & 6.31  & 11.61 & 1.32 & 5.15  & 5.54    \\ \hline
\begin{tabular}[c]{@{}c@{}}AGS (3DGS)\end{tabular}      & 1.50  & 4.54  & 4.91  & 0.39 & 2.70  & 2.81    \\ \hline
\begin{tabular}[c]{@{}c@{}}Orb-SLAM2 (Trad)\end{tabular}& 1.60  & 2.20  & 4.70  & 0.40 & 1.00  & 1.98    \\ \hline
\end{tabular}%
}
\vspace{-5pt}
\end{table}

\begin{figure}[!t]
\centering
\includegraphics[width=0.90\linewidth]{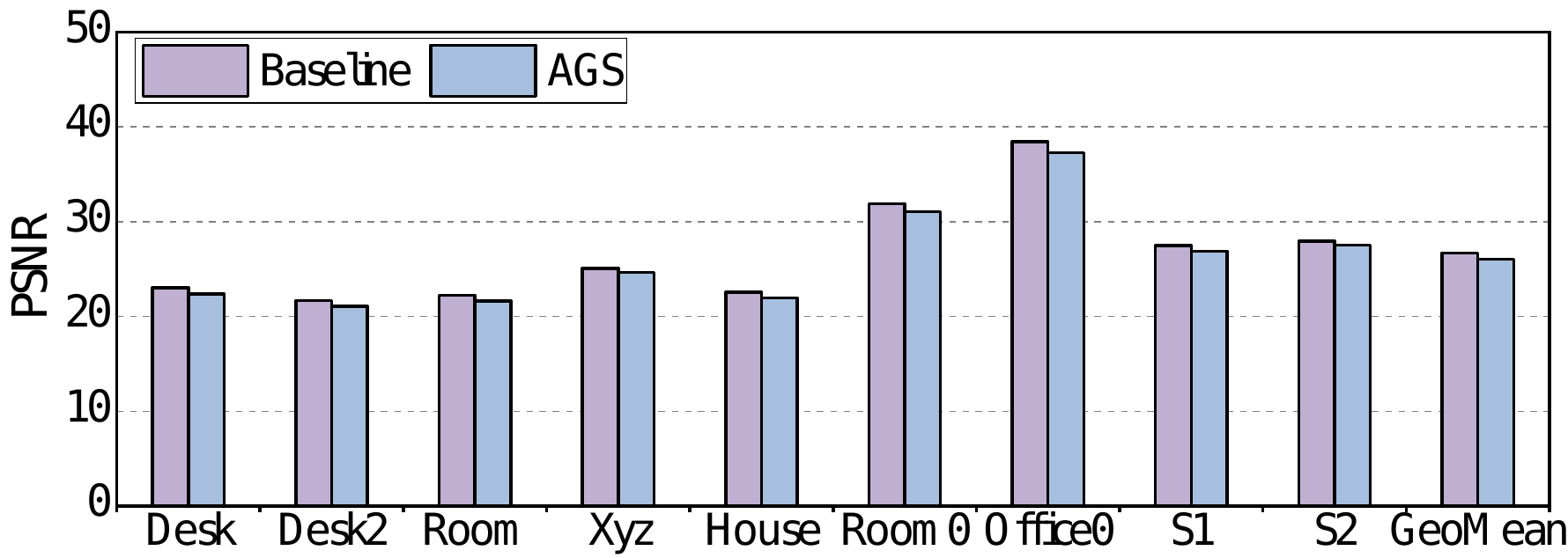}\vspace{-5pt}
\caption{PSNR$\uparrow$ of the baseline algorithm and AGS.}\vspace{-5pt}
\label{fig-PSNR}
\end{figure}

\subsection{Evaluation Settings}\label{ssect:setup}

\textbf{Datasets.} To evaluate the accuracy and performance improvement achieved by our proposed AGS architecture, we select three datasets: TUM-RGBD~\cite{sturm12iros}, Replica~\cite{replica19arxiv}, and Scannet++~\cite{yeshwanthliu2023scannetpp}, which are commonly used benchmarks for the evaluation of 3DGS-SLAM systems.

\textbf{Baselines.} Our evaluation compares the AGS architecture against two different classes of computing platforms. We choose NVIDIA Jetson AGX Xavier~\cite{Xavier} as a representative edge device, and A100~\cite{choquette2020nvidia} is selected as high-end server-level hardware. Moreover, we compare it with a state-of-the-art 3DGS accelerator, GSCore. For a fair comparison with the GPUs, we scale the number of computing cores to ensure the same area budget. To compare with GSCore, since it does not provide a feasible solution to accelerate 3DGS training, we combine the accelerated inference process of GSCore with the rest training process of the 3DGS-SLAM system on the GPUs for comparison.

\textbf{Architecture Implementation.}
To evaluate the performance of the AGS architecture, we develop a cycle-level simulator to mimic the hardware behavior of the AGS architecture. The simulator is integrated with Ramulator~\cite{kim2015ramulator} for DRAM timing. To measure the cycles, we run the datasets on GPUs using SplaTAM's official PyTorch implementation, collect point traces, and feed them into the simulator. Additionally, we implement the proposed AGS architecture in Verilog and synthesize it by Synopsys Design Compiler to get the chip area and total power under 28nm technology with a frequency of 500MHz. For SRAM-based on-chip buffers, we use CACTI 7~\cite{balasubramonian2017cacti} to model their area and power consumption under 32nm technology and scale them to 28nm using DeepScaleTool~\cite{sarangi2021deepscaletool}.

We evaluate two variants of AGS: AGS-Edge and AGS-Server. AGS-Edge is a design point with strict area and power constraints, which is a typical case for edge platforms. AGS-Server is a scaled-up architecture for high-end computing platforms. AGS-Edge contains a GS array of $16\times(4\times4)$ GPEs with smaller on-chip SRAM buffers, and AGS-Server contains a GS array of $32\times(4\times4)$ GPEs with larger on-chip SRAM buffers. We configure the off-chip memory of AGS-Edge as LPDDR4-3200 and AGS-Server as HBM2. We will discuss the hardware configurations and area of two variants of AGS in~\ref{ssect:exp-area}.

\subsection{Accuracy}\label{ssect:exp-performance}

Table~\ref{tab_Ate} and Fig.~\ref{fig-PSNR} compare the tracking accuracy and the mapping quality of the baseline implementation SplatAM and the proposed AGS algorithm, respectively. We assess these aspects using metrics including ATE RMSE (Absolute Trajectory Error Root Mean Square Error), commonly employed to measure the accuracy of tracking, and PSNR (Peak Signal-to-Noise Ratio) to quantify the rendering quality for mapping. 

For tracking, AGS achieves $1.97\times$ ATE RMSE improvement compared with SplatAM on the real-world TUM-RGBD dataset. Moreover, traditional SLAM systems demonstrate even higher precision owing to their precise geometric constraints. For instance, Orb-SLAM2 attains an average ATE RMSE of $1.98cm$, surpassing the performance of both 3DGS-based methods. Notably, tracking on synthetic datasets such as Replica presents significantly reduced challenge, with both AGS and the baseline achieving exceptional accuracy levels (e.g. $\leq0.5cm$). For mapping, the results show that AGS has an average of $2.36\%$ PSNR loss compared to SplatAM, whereas traditional SLAM methods lack the capability for photorealistic rendering, a key strength of 3DGS-based approaches (Table~\ref{tab_intro}). Thus, by synergizing 3DGS with traditional SLAM techniques, AGS effectively balances enhanced performance with minimal accuracy trade-offs.

We also evaluate the correctness of our prediction in Gaussian contribution-aware mapping by comparing the IDs of non-contributory Gaussians in the baseline algorithm with IDs of the ` predicted non-contributory Gaussians' generated by our proposed techniques. This comparison yields a false positive (FP) metric, where FP cases represent contributory Gaussians erroneously predicted as non-contributory. At just $5.7\%$, the average FP rate confirms the AGS algorithm's robustness, ensuring a negligible impact on the final rendering quality.

\begin{figure}[!t]
\centering
\includegraphics[width=0.90\linewidth]{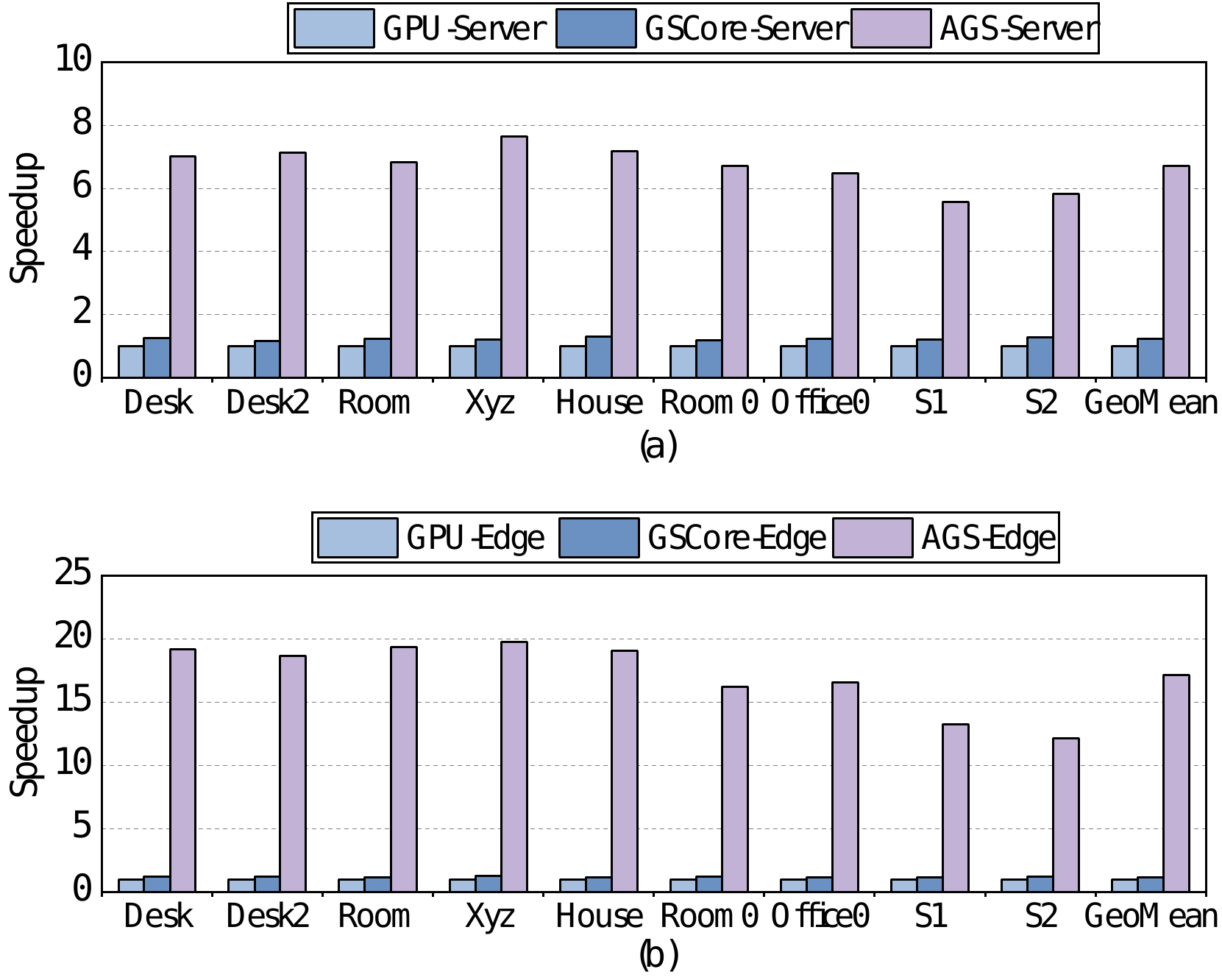}\vspace{-5pt}
\caption{Speedup of AGS-Server over A100 and GSCore (a); Speedup of AGS-Edge over AGX Xavier and GSCore (b).}\vspace{-5pt}
\label{fig-Speedup}
\end{figure}

\subsection{Performance}\label{ssect:exp-performance}

Fig.~\ref{fig-Speedup} showcases the performance of AGS over GPUs and GSCore. The results are normalized to GPUs.  As shown in Fig.~\ref{fig-Speedup}(a), AGS-Server averagely achieves $6.71\times$ and $5.41\times$ speedups over A100 and GSCore-Server. As illustrated in Fig.~\ref{fig-Speedup} (b), AGS-Edge delivers $17.12\times$ and $14.63\times$ speedups over AGX Xavier and GSCore-Edge. The reasons for being faster than GPUs include: 1) The movement-adaptive tracking saves redundant training iterations for tracking. 2) The Gaussian contribution-aware mapping avoids the redundant computation of non-contributory Gaussians for mapping. 3) We design a GPE scheduler, alpha buffers, and workload tables to balance the workloads between GPEs. Furthermore, we carefully allocate independent hardware resources and enable highly parallel operation of tracking and mapping. In contrast, GPUs encounter both redundant tracking iterations and non-contributory Gaussians and execute tracking and mapping sequentially. Moreover, AGS provides a higher speedup than GSCore for supporting the whole training process of 3DGS, while GSCore focuses only on accelerating the rendering process, which occupies a small amount of the execution time of 3DGS-SLAM systems.

\subsection{Area and Energy Efficiency}\label{ssect:exp-area}

\textbf{Hardware overhead and area.}
Table~\ref{tab_Area} provides an overview of the area and hardware configurations of AGS. AGS-Server and AGS-Edge have areas of 14.38$mm^2$ and 7.25$mm^2$, respectively. The pose tracking engine and the mapping engine are the two main components of GSA and occupy more than 90\% of the chip area. One notable factor contributing to the small area requirements of AGS is the collaboration between the FC detection engine and the CODEC. By leveraging the idle hardware resources of the CODEC, we effectively minimize the area overhead to capture frame covisibility. 

\begin{figure}[!t]
\centering
\includegraphics[width=0.90\linewidth]{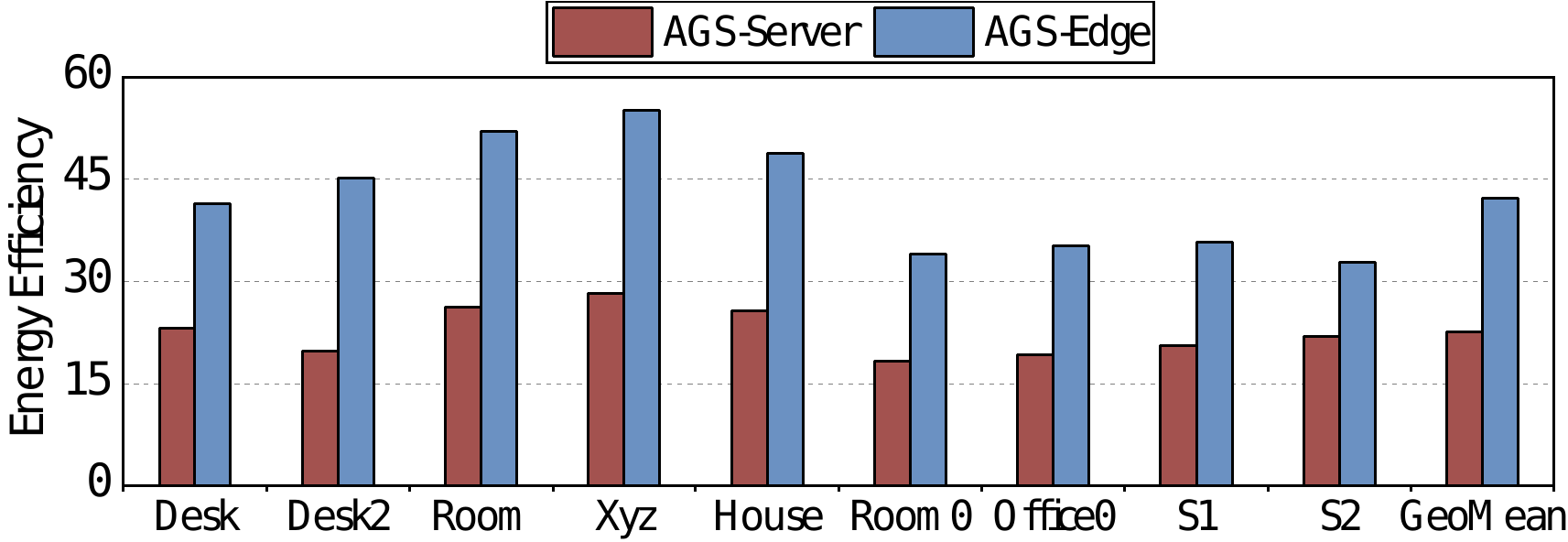}\vspace{-5pt}
\caption{Energy Efficiency of AGS over GPUs.}\vspace{-5pt}
\label{fig-Energy}
\end{figure}

\textbf{Energy efficiency.} 
Fig.~\ref{fig-Energy} shows the energy comparisons between AGS and GPUs, where the energy efficiency of AGS refers to the ratio of energy consumption by GPUs and AGS. As shown, AGS-Server surpasses the A100 GPU by $22.58\times$ in energy efficiency, and AGS-Edge achieves $42.28\times$ energy efficiency improvements over AGX Xavier. The significant savings in energy consumption mainly come from the reduction in excessive iterations during tracking, avoiding the computation of non-contributory Gaussians during mapping, and alleviating unbalanced workloads of the GPEs.

\begin{table}[!t]
\centering
\renewcommand\arraystretch{1.1}
\caption{Area of AGS.}
\vspace{-0pt}
\label{tab_Area}
\resizebox{\columnwidth}{!}{%
\begin{tabular}{llll}
\hline
Modules                                                                                 & Compomemt            & Remarks           & Area{[}$mm^2${]} \\ \hline
FC Detection Engine                                                                     & Adders and Comparators    & 8 Units+2 Units            & 0.01/0.01    \\ \hline
\multirow{4}{*}{\begin{tabular}[c]{@{}l@{}}Pose Tracking \\ Engine\end{tabular}}        & Systolic Array       & 2×(32×32)/4×(32×32) & 0.96/1.92    \\
                                                                                        & NN Buffer            & 32KB/64KB         & 0.09/0.13    \\
                                                                                        & GS Array (Light)     & 8×(4×4)/16×(4×4)    & 1.77/3.53    \\
                                                                                        & Gauss Buffer (Light) & 32KB/64KB         & 0.23/0.46    \\ \hline
\multirow{6}{*}{\begin{tabular}[c]{@{}l@{}}Mapping Engine\end{tabular}} & GS Logging Table     & 4KB/8KB           & 0.03/0.04    \\
                                                                                        & Update Unit          & 16 Units/32 Units & 0.13/0.25    \\
                                                                                        & GS Skipping Table    & 4KB/8KB           & 0.03/0.04    \\
                                                                                        & Comparison Unit      & 16 Units/32 Units & 0.01/0.01    \\
                                                                                        & GS Array             & 16×(4×4)/32×(4×4)   & 3.53/7.06    \\
                                                                                        & Gauss Buffer         & 64KB/128KB        & 0.46/0.93    \\ \hline
\textbf{Total Edge/Server}                                                                       &                      &                   &7.25/14.38              \\ \hline
\end{tabular}%
}
\vspace{-5pt}
\end{table}

\subsection{Ablation Study}\label{ssect:ablation}

\begin{figure}[!t]
\centering
\includegraphics[width=0.9\linewidth]{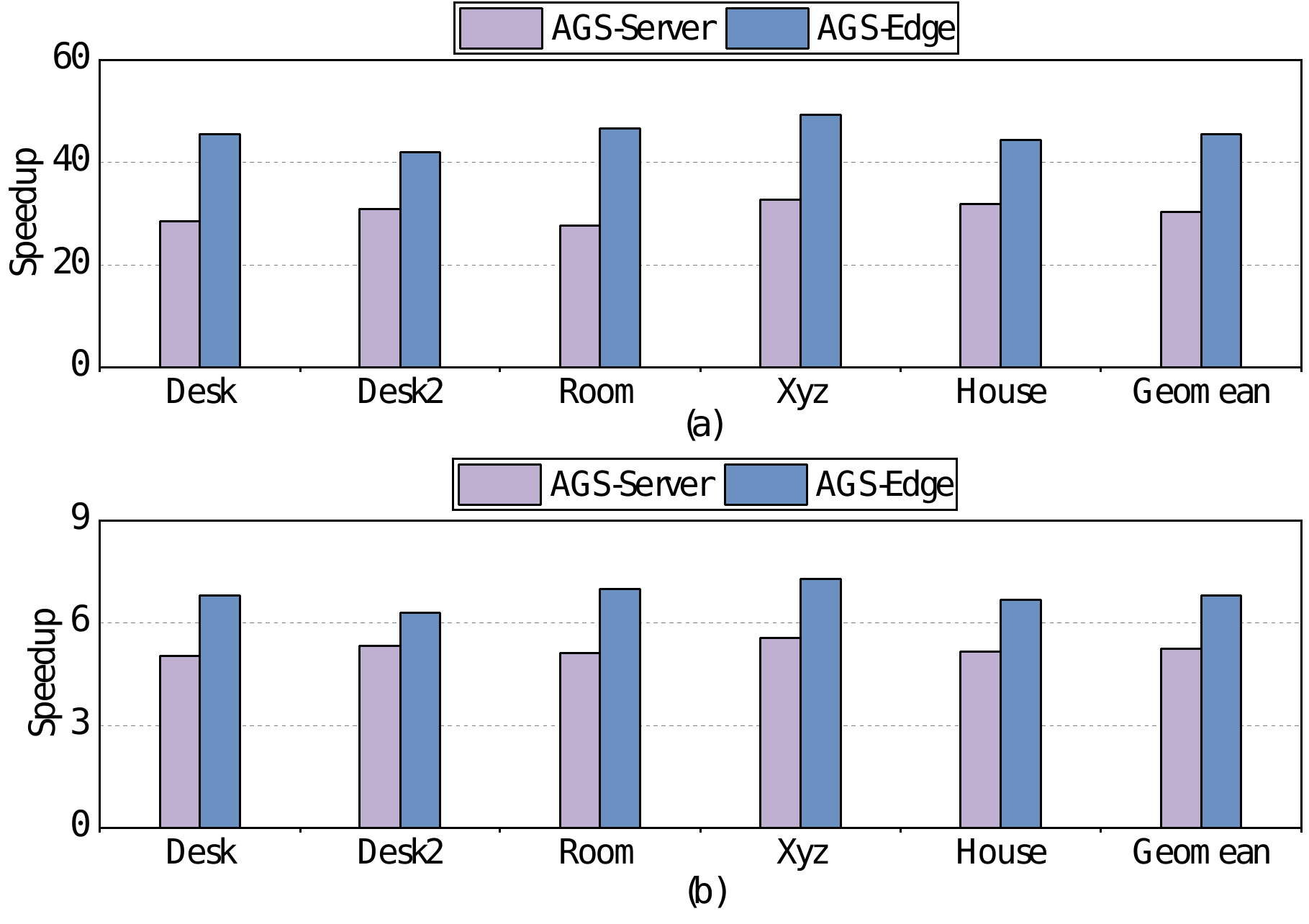}\vspace{-5pt}
\caption{Speedup on: (a) Tracking. (b) Mapping.}\vspace{-5pt}
\label{fig-Speedup_2stage}
\end{figure}

Fig.~\ref{fig-Speedup_2stage} compares the performance of AGS and GPUs for the two main tasks in 3DGS-SLAM: tracking and mapping. The results from TUM-RGBD show that AGS achieves a higher speedup on tracking than on mapping, attributed to the higher frame covisibility exhibited in tracking than mapping, which enables us to save a greater proportion of tracking's workload compared to mapping while maintaining a negligible quality loss. 

Also, we achieve a higher speedup rate on edge devices in comparison with server devices. This is because edge devices typically have lower bandwidth than server devices, which magnifies the memory inefficiency. Correspondingly, AGS happens to propose Gaussian contribution-aware mapping that avoids loading the non-contributory Gaussians and designs specialized memory units such as the GS logging/skipping table to reduce off-chip memory access, alleviating such inefficiencies.

\begin{figure}[!t]
\centering
\includegraphics[width=0.9\linewidth]{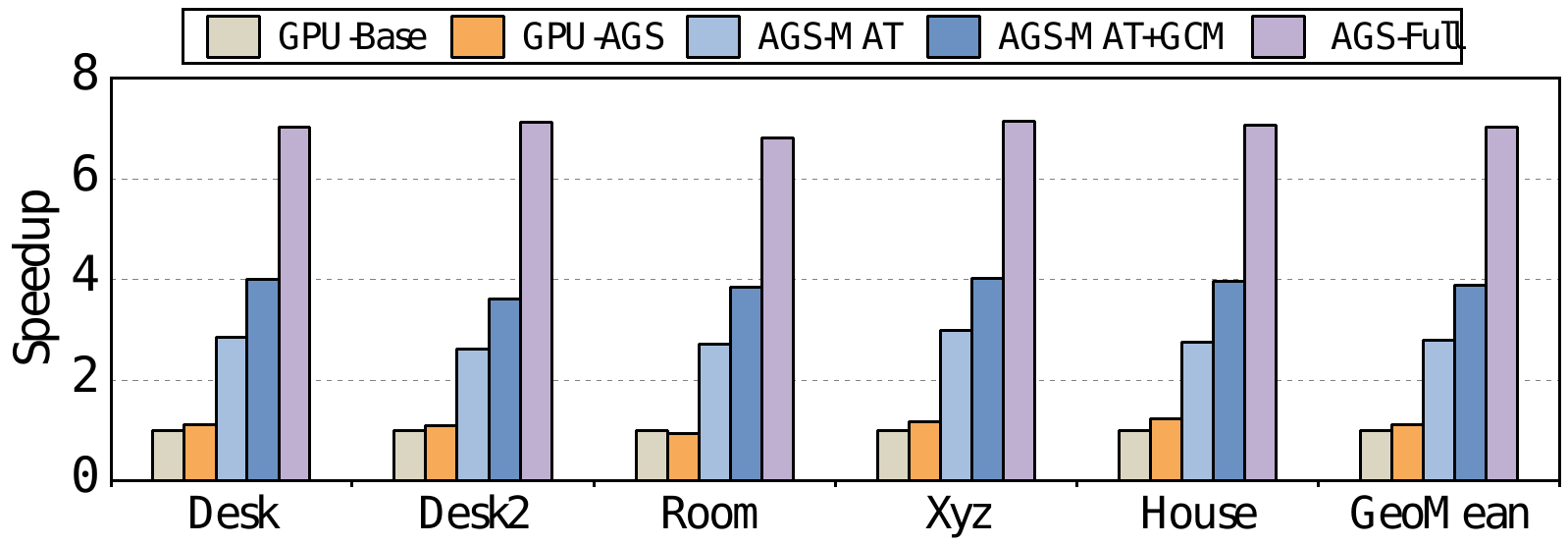}\vspace{-5pt}
\caption{Detailed analysis of contribution.}\vspace{-5pt}
\label{fig-Contribution}
\end{figure}

\textbf{Analysis of algorithm and architecture contributions.}  Fig.~\ref{fig-Contribution} depicts the acceleration gained from our algorithm and architectural designs. Compared to running the baseline algorithm (GPU-Base), implementing the AGS algorithm on GPU only provides a $1.12\times$ speedup (GPU-AGS). This is because: 1) Without utilizing the CODEC, GPU conducts the frame covisibility detection and the SLAM pipeline in a serial manner. 2) GPU struggles to efficiently perform the extra memory operations introduced by Gaussian contribution-aware mapping, including recording and requesting the Gaussian contribution information. 

\begin{table}[!t]
\centering
\renewcommand\arraystretch{1.3}
\caption{Comparison with directly integrating SplatAM with Droid-SLAM (PSNR, higher value indicates finer quality).}
\label{tab_Direct}
\setlength{\arrayrulewidth}{0.85pt}
\resizebox{0.9\columnwidth}{!}{%
\begin{tabular}{ccccccc}
\hline
Benchmark        & Desk   & Desk2   & Room    & Xyz    & House  & Geomean   \\ \hline
AGS              & 21.4   & 20.07   & 20.64   & 24.67  & 20.98  & 21.55     \\ \hline
Droid+SplatAM    & 20.65  & 19.13   & 20.38   & 24.03  & 20.16  & 20.87     \\ \hline
\end{tabular}%
}
\vspace{-0pt}
\end{table}

\begin{figure}[!t]
\centering
\includegraphics[width=0.9\linewidth]{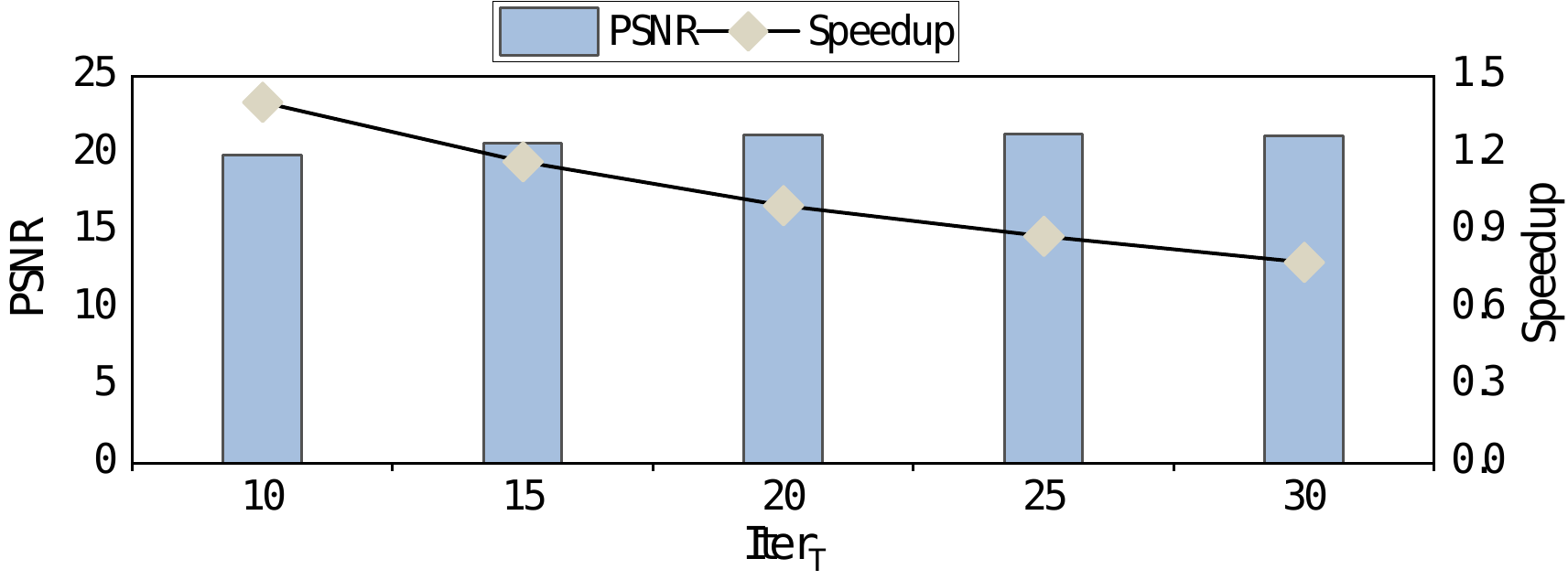}\vspace{-5pt}
\caption{Analysis of the threshold $iter_{T}$.}\vspace{-5pt}
\label{fig-Exp_iter}
\end{figure}

\begin{figure}[!t]
\centering
\includegraphics[width=0.9\linewidth]{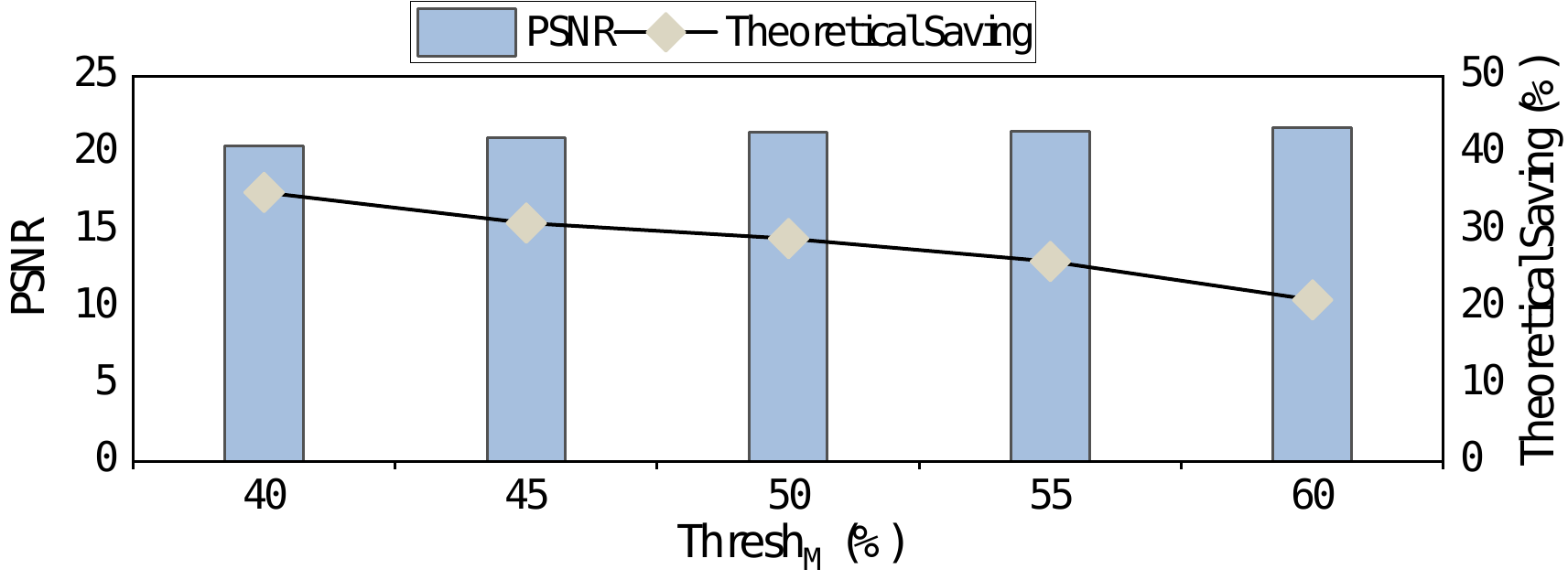}\vspace{-5pt}
\caption{Analysis of the threshold $Tresh_{M}$.}\vspace{-5pt}
\label{fig-Exp_M}
\end{figure}

\begin{figure}[!t]
\centering
\includegraphics[width=0.9\linewidth]{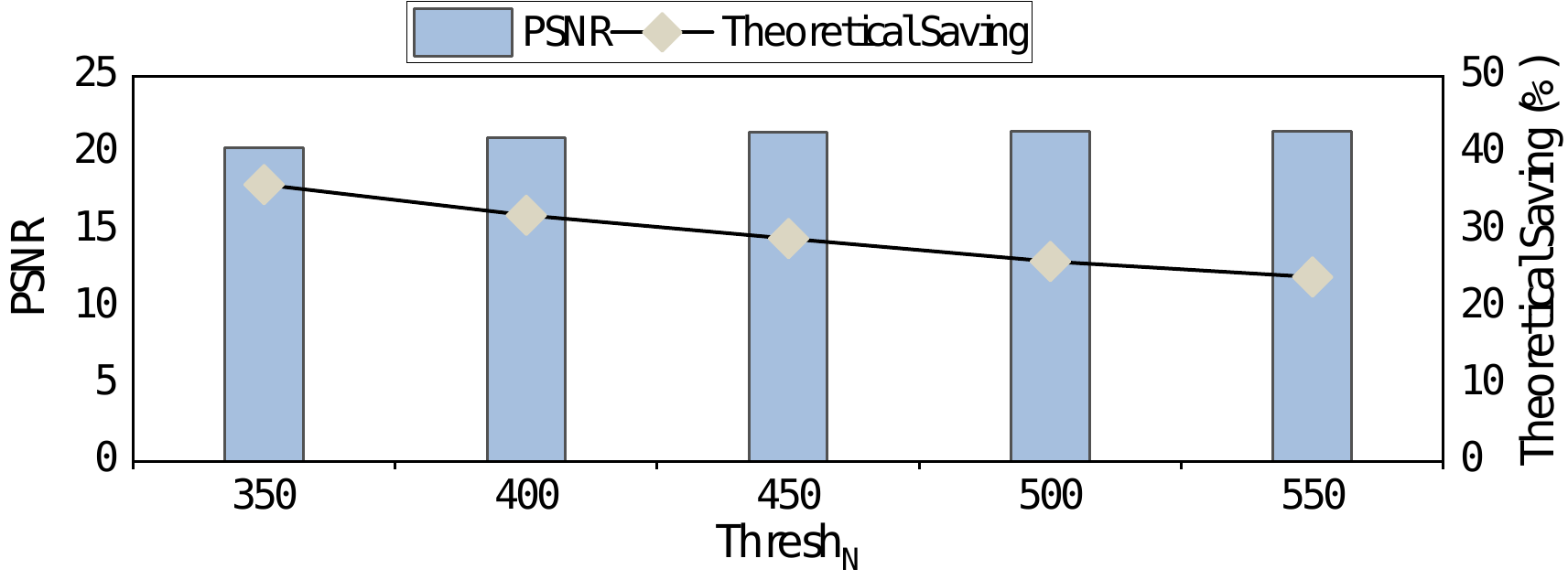}\vspace{-5pt}
\caption{Analysis of the threshold $Tresh_{N}$.}\vspace{-5pt}
\label{fig-Exp_N}
\end{figure}

Compared to GPU-AGS, the dedicated hardware design for movement-adaptive tracking (AGS-MAT) and Gaussian contribution-aware mapping (AGS-MAT+ GCM) further enhances the performance by $2.51\times$ and $1.42\times$. Moreover, the proposed GS array (denoted as AGS-Full) brings an extra $1.79\times$ speedup over AGS-MAT+GCM. This improvement stems from the collaboration between the GPEs and the GPE scheduler, which jointly optimize the 3DGS pipeline, thereby improving hardware utilization.

\textbf{Analysis of directly integrating SplatAM with Droid-SLAM.}
As shown in table~\ref{tab_Direct}, we compare the PSNR achieved by AGS with directly integrating SplatAM with Droid-SLAM. Results show that the direct approach (Droid+SplatAM) degrades mapping quality (PSNR) to 20.87 dB, where AGS achieves 21.55 dB. This is because the camera pose generated from Droid-SLAM is not inherently compatible with the scene reconstructed by 3DGS. This validates the critical role of the proposed fine-grained tracking refinement in maintaining the mapping accuracy.

\begin{figure}[!t]
\centering
\includegraphics[width=0.9\linewidth]{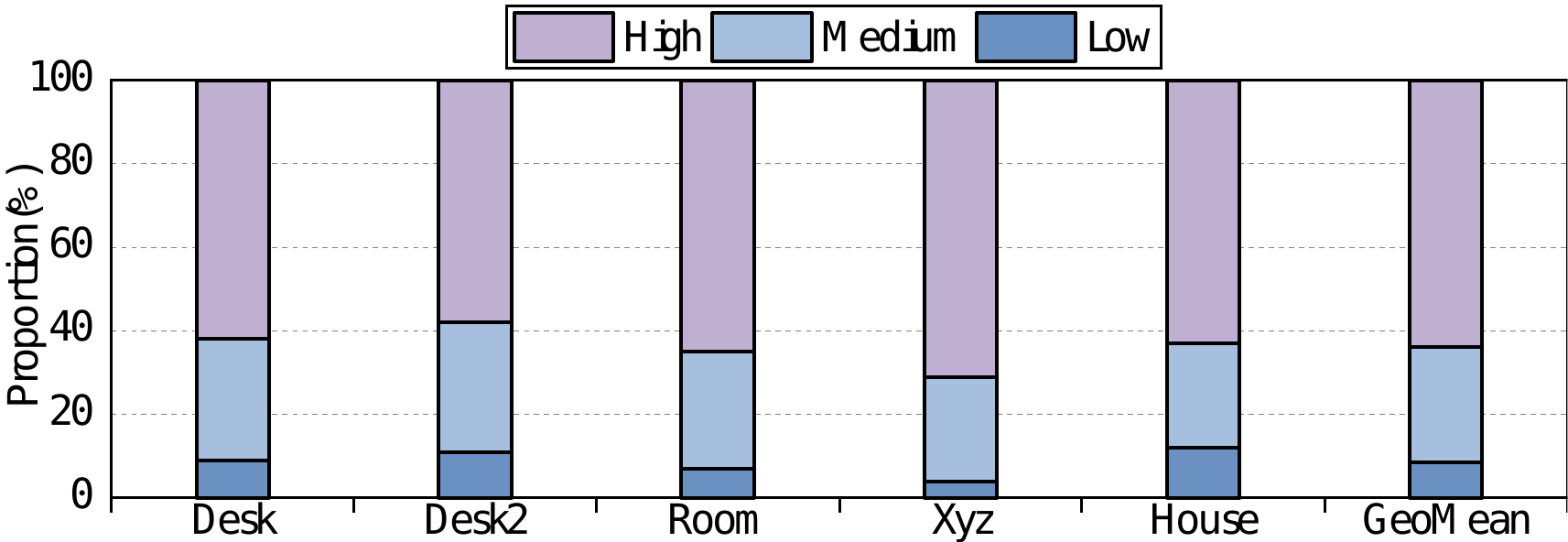}\vspace{-0pt}
\caption{The proportion of adjacent frames exhibiting varying covisibility levels.}\vspace{-0pt}
\label{fig-FC}
\end{figure}

\begin{figure}[!t]
\centering
\includegraphics[width=0.9\linewidth]{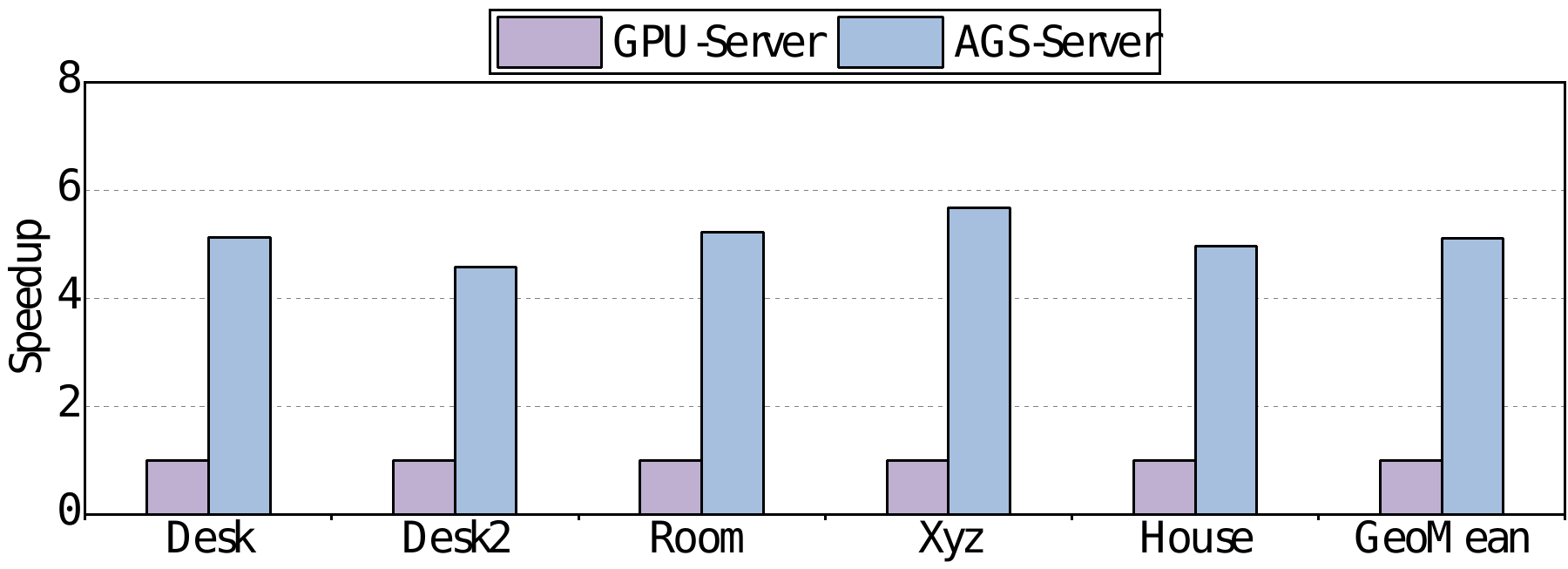}\vspace{-0pt}
\caption{Performance improvement of AGS on Gaussian-SLAM.}\vspace{-0pt}
\label{fig-Gaussian-SLAM}
\end{figure}

\subsection{Discussion}\label{ssect:exploration}

\textbf{Sensitivity study.} The number of refinement tracking iterations $Iter_{T}$, the mapping threshold to define key frames and non-key frames $Thresh_{M}$, and the threshold to define the contribution of Gaussians $Thresh_{N}$ affect the performance of AGS. In this section, we try to determine the three hyperparameters by striking a balance between performance and accuracy. The experiments are conducted on the Desk scene. 

Specifically, Fig.~\ref{fig-Exp_iter} studies the impact of the value $Iter_{T}$. When $Thresh_{T}$ increases, fewer iterations are reduced by the proposed movement-adaptive tracking algorithm, leading to lower performance gain in exchange for higher rendering quality. As a result, we set $Iter_{T}$ to $20$. Fig.~\ref{fig-Exp_M} depicts that when $Thresh_{M}$ increases, fewer frames are designated as non-key frames, resulting in reduced theoretical savings from selective mapping but enhanced rendering quality, which indicates us to set $Thresh_{M}$ as $50\%$. As shown in Fig.~\ref{fig-Exp_N}, when $Thresh_{N}$ increases, fewer Gaussians are regarded as non-contributory Gaussians and will not be skipped by selective mapping for the non-key frames. As a result, we set $Thresh_{N}$ as $450$ to achieve a balance between quality and performance.

\textbf{Analysis on frame covisibility.} As AGS's performance depends on the frame covisibility (FC), we evaluate the FC across adjacent frames on the TUM-RGBD dataset. Specifically, for each incoming frame, we compute the FC with its previous frame and record the proportion of different covisibility levels. As shown in Fig.~\ref{fig-FC}, $63.8\%$ of the adjacent frames exhibits high FC for tracking, which is the main source of performance improvement due to movement-adaptive tracking. This demonstrates that the SLAM pipeline typically provides sufficiently high FC that allows AGS to save redundant computations.

\textbf{Generality of AGS.}To validate the generality of AGS in different 3DGS-SLAM backbones, we implement Gaussian-SLAM, another representative 3DGS-SLAM algorithm on AGS (AGS-Server) and compare its performance with GPU-accelerated Gaussian-SLAM (GPU-Server). For AGS-Server, we integrate the Gaussian-SLAM algorithm into our AGS simulator to evaluate the performance gains achieved by using AGS for Gaussian contribution-aware mapping and hardware scheduling. For GPU-Server, we leverage the implementation of Gaussian-SLAM~\cite{yugay2023gaussian} and test its performance on A100. Fig.~\ref{fig-Gaussian-SLAM} demonstrates that AGS enhances Gaussian-SLAM with a $5.11\times$ speedup, proving that AGS is a versatile solution applicable to emerging 3DGS-SLAM algorithms.

%% file: Related.tex
\section{Related Work}
\subsection{Algorithm optimization}

\textbf{3DGS-SLAM Models.} Significant work has been done on integrating 3DGS into the SLAM framework~\cite{matsuki2024gaussian, keetha2024splatam, yugay2023gaussian, yan2024gs}. SplaTAM~\cite{keetha2024splatam} enables high-quality image rendering with key steps including pose tracking, Gaussian densification, and map update. Specially, it utilizes a silhouette mask to capture the presence of scene density. GS-SLAM~\cite{yan2024gs} introduces an adaptive expansion strategy, dynamically adding Gaussians to represent newly observed scene geometry while removing redundant ones that deviate significantly from the ground truth. Gaussian-SLAM~\cite{yugay2023gaussian} introduces an extra scale regularization loss and creates separate sub-maps to prevent catastrophic forgetting for computational efficiency. 

\textbf{Gaussian contribution-aware algorithms.} Algorithmic approaches have also been proposed to address the redundancy caused by non-contributory Gaussians in 3DGS. To represent scenes with a constrained number of Gaussians, Mini-Splatting~\cite{fang2024mini} introduces strategies for densification including blur split and depth reinitialization, which optimize Gaussian spatial distribution to enhance both computational efficiency and memory usage. MetaSapiens~\cite{lin2025metasapiens} implements a real-time Point-Based Neural Rendering (PBNR) system by presenting an efficiency-aware Gaussian pruning technique. It also introduces a Foveated Rendering (FR) method for PBNR, leveraging humans’ low visual acuity in peripheral regions to relax rendering quality for acceleration.

\subsection{Leveraging Inter-Frame Similarities}\label{ssect:prior}

Recent years have seen explorations of leveraging inter-frame similarities to accelerate computer vision algorithms, with many of them targeting CNN inference tasks~\cite{buckler2018eva2, zhu2018euphrates, gan2020low}. EVA$^2$~\cite{buckler2018eva2} proposes an activation motion compensation algorithm that detects changes in the visual input and reuses the previously-computed activation to save redundant executions. Similarly, Euphrates~\cite{zhu2018euphrates} leverages motion information extracted from the Image Signal Processor (ISP) to predict the changes in pixel data, thereby reducing the number of computations during CNN inference. 

Other works also utilize inter-frame similarities for point cloud analytics~\cite{kammerl2012real}, vision systems~\cite{zhao2021holoar, zhao2020deja, feng2019asv}, and NeRF rendering~\cite{feng2024cicero}. Among them, Cicero~\cite{feng2024cicero} represents the most closely related accelerator to the proposed AGS. The key innovation of Cicero lies in its exploitation of ray similarity, which enables direct reusage of the rendering results of key rays to bypass computations for similar rays. Unfortunately, Cicero remains limited to ray-level optimization and overlooks the opportunity at the fine-grained per-Gaussian level, whereas AGS excavates the redundancy brought by non-contributive Gaussians. On the hardware level, the performance bottleneck of NeRF lies in the hash encoding and Multilayer Perceptron (MLP) operations, presenting distinct architectural challenges compared to the 3DGS pipeline. Therefore, the hardware design of Cicero is infeasible to accelerate 3DGS-based algorithms.

\subsection{Hardware Acceleration of 3DGS}

Additionally, we introduce hardware accelerators that improve the performance of 3DGS. GSCore~\cite{lee2024gscore} is the first design to accelerate the inference process of 3DGS. It observes that not all Gaussians assigned to related tiles are eventually useful. To address this, GSCore orchestrates the rendering pipeline of 3DGS with the Gaussian-shape-aware intersection test, hierarchical sorting, and sub-tile skipping techniques to predict and skip the preprocessing for unused Gaussians, thereby reducing the workload of rendering. 

Iris~\cite{song2025iris} identifies the bottleneck of 3DGS rendering, including its massive external memory access and huge computational requirements. To enable interactive photorealistic rendering, Iris first proposes coplanar Gaussian cluster and spatial-temporal Gaussian order reuse to eliminate redundant Gaussian fetching and sorting. It also designs a reconfigurable multiply-accumulator array and an error direction cache to realize algorithmic operations with high energy efficiency and throughput.

%% file: Conclusion.tex
\section{Conclusion}

This paper introduces AGS, a specialized accelerator poised to facilitate an efficient training pipeline of 3DGS-SLAM. The key idea is to utilize the frame covisibility between frames extracted from video CODEC to overcome the inefficiencies of both tracking and mapping. Extensive experiments show that AGS can deliver satisfactory performance gain with trivial accuracy loss.

\section{Acknowledgements}
We thank the anonymous reviewers and our shepherd Venkat Arun for their valuable feedback. This work is partly supported by the National Natural Science Foundation of China (Grant No. 62202288). Zhuoran Song is the corresponding author.